\documentclass[12pt,preprint]{aastex}

\shortauthors{MANCUSO ET AL.}
\shorttitle{GALAXY EVOLUTION IN THE RADIO BAND}

\begin{document}

\title{Galaxy Evolution in the Radio Band:\\ The Role of Starforming Galaxies and Active Galactic Nuclei}
\author{C. Mancuso\altaffilmark{1,2,3}, A. Lapi\altaffilmark{2,3,4}, I. Prandoni\altaffilmark{1}, I. Obi\altaffilmark{2}, J. Gonzalez-Nuevo\altaffilmark{5}, F. Perrotta\altaffilmark{2,4}, A. Bressan\altaffilmark{2,3,4}, A. Celotti\altaffilmark{2,3,6}, L. Danese\altaffilmark{2,3,4}}\altaffiltext{1}{INAF-IRA, Via P. Gobetti 101, 40129 Bologna, Italy}\altaffiltext{2}{SISSA, Via Bonomea 265, 34136 Trieste, Italy}\altaffiltext{3}{INFN-Sezione di Trieste, via Valerio 2, 34127 Trieste, Italy}\altaffiltext{4}{INAF-Osservatorio Astronomico di Trieste, via Tiepolo 11, 34131 Trieste, Italy}\altaffiltext{5}{Departamento de Fisica, Universidad de Oviedo, C. Calvo Sotelo s/n, E-33007 Oviedo, Spain}\altaffiltext{6}{INAF-Osservatorio Astronomico di Brera, via Bianchi 46, 23807 Merate, Italy}

\begin{abstract}
We investigate the astrophysics of radio-emitting star-forming galaxies and active galactic nuclei (AGNs), and elucidate their statistical properties in the radio band including luminosity functions, redshift distributions, and number counts at sub-mJy flux levels, that will be crucially probed by next-generation radio continuum surveys. Specifically, we exploit the model-independent approach by Mancuso et al. (2016a,b) to compute the star formation rate functions, the AGN duty cycles and the conditional probability of a star-forming galaxy to host an AGN with given bolometric luminosity. Coupling these ingredients with the radio emission properties associated to star formation and nuclear activity, we compute relevant statistics at different radio frequencies, and disentangle the relative contribution of star-forming galaxies and AGNs in different radio luminosity, radio flux, and redshift ranges. Finally, we highlight that radio-emitting star-forming galaxies and AGNs are expected to host supermassive black holes accreting with different Eddington ratio distributions, and to occupy different loci in the galaxy main sequence diagrams. These specific predictions are consistent with current datasets, but need to be tested with larger statistics via future radio data with multi-band coverage on wide areas, as it will become routinely achievable with the advent of the SKA and its precursors.
\end{abstract}


\setcounter{footnote}{0}

\section{Introduction}\label{sec|intro}

Recent wide-area far-IR/(sub-)mm surveys conducted by Herschel, ASTE/AzTEC, APEX/ LABOCA, JCMT/SCUBA2, and ALMA-SPT (e.g., Gruppioni et al. 2013; 2015; Lapi et al. 2011; Weiss et al. 2013; Strandet et al. 2016; Koprowski et al. 2014, 2016), in many instances eased by gravitational lensing from foreground objects (Negrello et al. 2014, 2017; Nayyeri et al. 2016), have revealed an abundant population of dusty star-forming galaxies (SFG) at high redshift $z\ga 1$, responsible for the bulk of the cosmic star formation history (Mancuso et al. 2016a; Lapi et al. 2017). Continuity equation arguments have undoubtedly demonstrated that these galaxies constitute the high-redshift progenitors of local ellipticals (Aversa et al. 2015; Mancuso et al. 2016a,b), and as such the future hosts of the most massive black holes (BHs) in the Universe. At redshifts $z\ga 1$ the growth of the central BH in the early stages of a massive galaxy's evolution has been caught in the act by X-ray and mid-IR followup observations of far-IR/sub-mm selected galaxies (e.g., Mullaney et al. 2012; Johnson et al. 2013; Wang et al. 2013; Delvecchio et al. 2015; Rodighiero et al. 2015), while its quenching effect on the star formation activity in the late stages has been indirectly revealed by far-IR follow-up observations of X-ray selected active galactic nuclei (AGNs; e.g., Page et al. 2012; Barger et al. 2015; Stanley et al. 2015; Harrison et al. 2016) or optically selected quasars (e.g., Omont et al. 2003; Mor et al. 2012; Xu et al. 2015; Netzer et al. 2016; Harris et al. 2016). At redshifts $z\la 1$, on the other hand, evidences of AGN-induced star formation have been found, especially in association with jetted emission from the nucleus (e.g., Kalfountzou et al. 2014; Rosario et al. 2015).

The study of high-redshift SFGs is of paramount importance to address the issue of coevolution between galaxies and supermassive BHs (e.g., Alexander \& Hickox 2012). However, current sensitivity limits of far-IR/sub-mm instruments do not allow to characterize the statistical properties of the SFG population at redshift appreciably larger than $z\ga 3$; in this perspective a new observational window, unbiased with respect to dust obscuration, will be provided by the upcoming ultradeep radio continuum surveys planned on SKA and its precursors (see Prandoni \& Seymour 2015; Norris et al. 2013).

Indeed, while radio-loud (RL) AGN dominate the radio sky all the way down to the sub-mJy regime  (78\% at $S_{\rm 1.4 GHz}\ga 0.5$ mJy, see Mignano et al. 2008), SFGs  gradually emerge at sub-mJy flux densities and eventually become the most relevant population below $S_{\rm 1.4 GHz}\la 100\, \mu$Jy (e.g., Simpson et al. 2006; Seymour et al. 2008; Smolcic et al. 2008). This also corresponds to a gradual change of the physical processes probed by deep radio surveys. In most RL AGNs the radio emission is associated with large-scale relativistic jets powered by BHs, hosted at the center of low redshift $z\la 1$ massive ellipticals  (e.g., Heckman \& Best 2014; Kellermann et al. 2016; Padovani 2016 and references therein). In SFGs, on the other hand, we mainly probe synchrotron (and free-free) radio emission associated to star-forming regions in the host galaxy (e.g., Condon 1992).

This relatively simple scenario has recently become more complex, as a third population has been detected at sub-mJy fluxes. In fact, moving toward fluxes $S_{1.4\,GHz}\la 0.1$ mJy (where SFGs already dominate) the RL AGN population is progressively outnumbered by the so-called radio-quiet (RQ) AGNs, i.e. galaxies showing clear signatures of AGN activity at non-radio wavelengths (e.g., X-ray, mid-IR, and optical) but with no signs of large-scale radio jets, and featuring much weaker radio emission than RL systems (e.g., Kellermann et al. 2016; Padovani 2016).

Two important issues regarding RQ AGNs are still hotly debated. First, the processes responsible for the radio emission in RQ AGNs are not well understood yet. Observational indications both of a nuclear and of a star formation origin have been reported. On the one hand, Padovani et al. (2015) and Bonzini et al. (2015) have shown that RQ AGNs feature infrared-to-radio flux ratios, evolving radio luminosity functions, host galaxy colors, optical morphologies and stellar masses similar to those of star-forming systems, suggesting that in RQ AGNs the radio emission is on the average dominated by star formation (see also Kimball et al. 2011; Condon et al. 2013; Kellermann et al. 2016). On the other hand, White et al. (2015, 2017) have argued that RQ AGNs show a radio luminosity exceeding that of SFGs of similar stellar masses. Note, however, that these distinct findings may be partly attributed to the different luminosity and redshift ranges probed by the above studies.

High-resolution (milli-arcsec) radio observations by Jackson et al. (2015), Maini et al. (2016) and Herrera-Ruiz et al. (2016) have revealed that RQ AGNs can contain nuclear radio cores significantly contributing to the total radio emission.  On a larger statistical ground, deep sub-arcsec resolution radio observations of the GOODS-N field have revealed that RQ AGNs are preferentially associated to more compact radio emission than star-forming galaxies (Guidetti et al. 2017). In addition, Zakamska et al. (2016) have shown that radio luminosities in RQ quasars exceed by an order of magnitude the ones expected from star formation. A plausible scenario is that star formation and nuclear radio emissions coexist in RQ AGNs, though it is still unclear which one dominates, at least in a statistical sense. Indeed, in the local Universe (at $z\la 0.5$) it is found that both AGN and star-formation processes can contribute to the total radio emission in RQ AGNs (e.g., Seyfert $2$ galaxies; Roy et al. 1998), and composite AGN and star-forming systems are common at medium to high redshift $z\ga 1-2$ (see Daddi et al. 2007; Del Moro et al. 2013; Rees et al. 2016).

Second, early evidence of a dichotomy between RL and RQ sources has been challenged, and is still controversial. On the one hand, Kellermann et al. (1989), Miller et al. (1990), Ivezic et al. (2002) have suggested a neat dichotomy in the radio-loudness distribution of such objects. In the same vein, Bonzini et al. (2015) and Padovani et al. (2015) have recently claimed that RQ and RL AGNs constitute totally distinct populations, characterized by very different evolutions, luminosity functions, and Eddington ratios. On the other hand, Lacy et al. (2001), Cirasuolo et al. (2003), Balokovich et al. (2012),  Bonchi et al. (2013) found continuous radio-loudness distributions with marginal evidence for a dichotomy; the same conclusion was also reached by Barvainis et al. (2005) basing on variability arguments.

Crucial issues that still need to be addressed are the following: is the radio luminosity function of non-RL systems dominated by star formation or nuclear emission, and in which luminosity ranges? is the amount of star formation in AGN hosts sufficient to explain the radio counts associated to sub-mJy radio sources, or is a substantial nuclear contribution necessary? is there a physical dichotomy between RL and RQ sources, or the two population smoothly connect, at least in a statistical sense? Next-generation radio surveys with SKA and its precursors will allow us to fully probe the SFG and RQ AGN populations, reaching unprecedented sensitivities (sub-$\mu$Jy) for the deepest fields, and/or providing wide-area samples at the depth (around $\mu$Jy) now achieved only by the deepest (and tiny) radio surveys. In combination with deep multi-wavelength information, this will provide an unbiased view of star formation, nuclear activity and of their interplay across cosmic times.

In this paper we tackle such issues by providing a novel view on the astrophysics and on the statistical properties of SFGs and AGNs in the radio band. To this purpose we take up the model-independent approach by Mancuso et al. (2016a,b), based on two basic ingredients: (i) the redshift-dependent SFR functions inferred from the latest UV/far-IR data; (ii) deterministic tracks for the coevolution of star formation and BH accretion in an individual galaxy, gauged on a wealth of multi-wavelength observations. We exploit such ingredients to compute the AGN duty cycle and probability of a SFG to host an AGN, so mapping the SFR functions into the observed bolometric AGN luminosity functions. Coupling these results with the radio emission properties associated to star formation and nuclear activity, we compute relevant statistics (like luminosity functions, redshift distributions and counts) at different radio frequencies, to disentangle the role of SFGs, RQ and RL AGNs in different luminosity/flux ranges.

Our predictions are compared against state-of-the-art deep radio surveys in extra-galactic fields where dense multi-band coverage is available, allowing a reliable classification of the radio sources. In particular we exploit one of the largest deep radio samples available to date: a $1.4$ GHz mosaic covering more than $6$ deg$^2$ in the Lockman Hole (LH) region down to an rms sensitivity of $11\,\mu$Jy per beam (Prandoni et al. 2017). This dataset, together with the wide multi-band data available in the LH region yields one of the most reliable source counts determination in the range $0.1-1$ mJy, and a very robust statistical decomposition of the relative contributions from SFGs, RQ and RL AGNs in this flux range (see Prandoni et al. 2017 for more details).

The plan of the paper is as follows. In \S~\ref{sec|basics} we describe the basic ingredients of our analysis: the SFR functions, the mapping of these into AGN luminosity functions, and the associated probability of occupation for AGNs in host SFGs; in \S~\ref{sec|radio} we discuss the radio emission properties from star formation and nuclear activity, and compute the related statistics in the radio band; in \S~\ref{sec|results} we present and discuss our results; in \S~\ref{sec|summary} we summarize our findings.

Throughout this work we adopt the standard flat cosmology (Planck Collaboration XIII 2016) with round parameter values: matter density $\Omega_M = 0.32$, baryon density $\Omega_b = 0.05$, Hubble constant $H_0 = 100\, h$ km s$^{-1}$ Mpc$^{-1}$ with $h = 0.67$, and mass variance $\sigma_8 = 0.83$ on a scale of $8\, h^{-1}$ Mpc. Stellar masses and SFR (or luminosities) of galaxies are evaluated assuming the Chabrier's (2003) initial mass function (IMF).

\section{Basic ingredients}\label{sec|basics}

Our analysis relies on two basic ingredients: (i) a model-independent determination of the SFR functions at different redshifts; (ii) deterministic evolutionary tracks for the history of star formation and BH accretion in an individual galaxy. In this section we briefly recall the basic notions relevant for the analysis of galaxy statistics in the radio band, deferring the reader to the papers by Mancuso et al. (2016a,b) for a detailed description.

\subsection{SFR functions}\label{sec|SFR_func}

First ingredient is constituted by the global SFR function ${\rm d}N/{\rm d}\log \dot M_\star$, namely the number density of galaxies per logarithmic bin of SFR $[\log \dot M_\star,\log\dot M_\star+{\rm d}\log\dot M_\star]$ at given redshift $z$. This has been accurately determined by Mancuso et al. (2016a,b) by exploiting the most recent determinations of the evolving galaxy luminosity functions from far-IR and UV data.

\begin{figure*}
\epsscale{1}\plotone{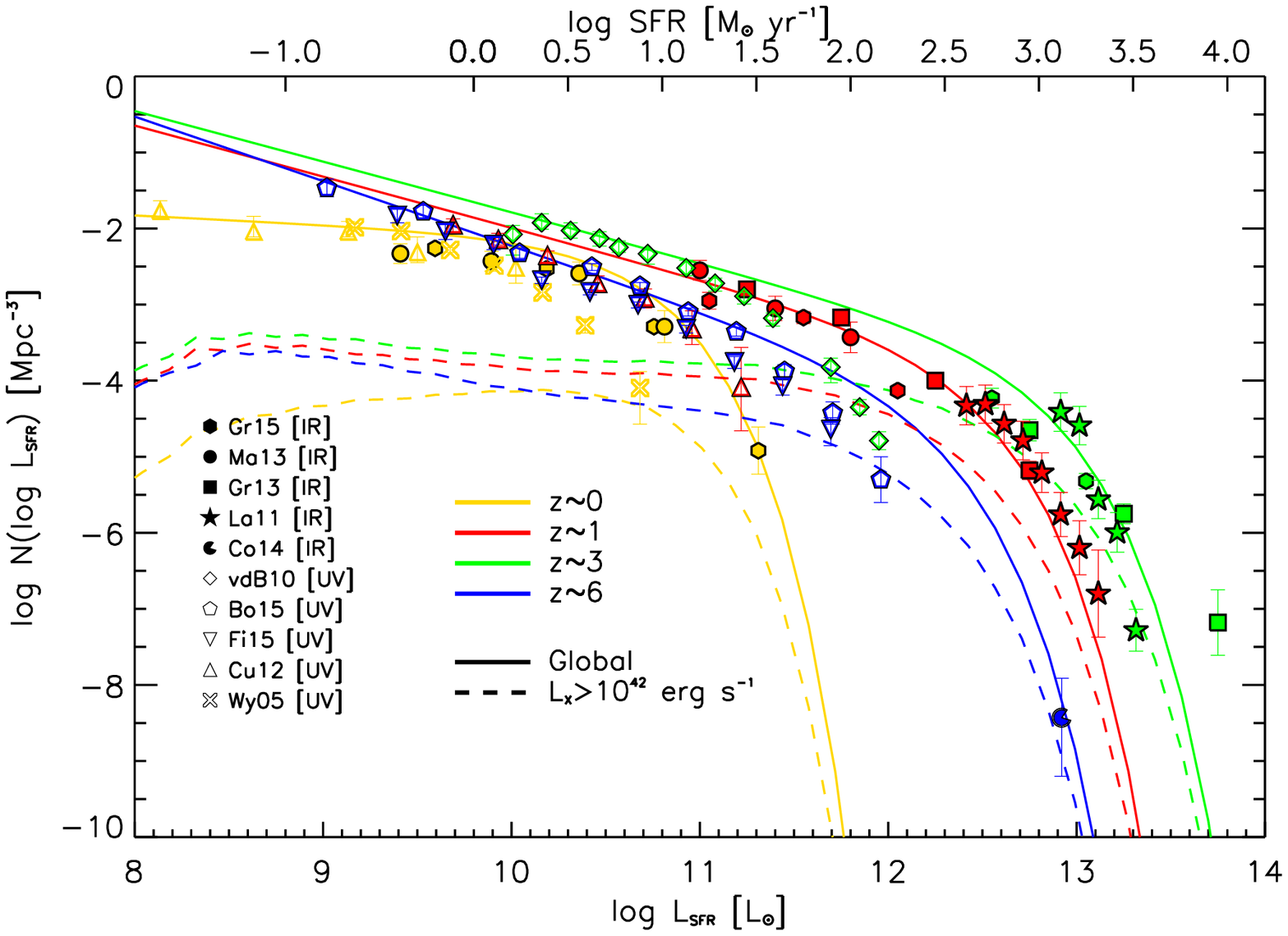}\caption{The SFR functions at redshifts $z=0$ (yellow), $1$ (red), $3$ (green) and $6$ (blue) determined according to the procedure by Mancuso et al. (2016a,b). Solid lines refer to the global SFR function based on (dust-corrected) UV plus far-IR measurements, while dashed lines are the SFR functions of galaxies hosting an AGN with X-ray luminosity larger than $10^{42}$ erg s$^{-1}$. UV data (open symbols) are from van der Burg et al. (2010; diamonds), Bouwens et al. (2015; pentagons) and Finkelstein et al. (2015; inverse triangles), Cucciati et al. (2012; triangles), and Wyder et al. (2005; crosses); far-IR data from Gruppioni et al. (2015; hexagons), Magnelli et al. (2013; circles), Gruppioni et al. (2013; squares), Lapi et al. (2011; stars), and Cooray et al. (2014; pacmans).}\label{fig|SFR_func}
\end{figure*}

In a nutshell, UV data have been dust-corrected according to the local empirical relation between the UV slope $\beta_{\rm UV}$ and the IR-to-UV luminosity ratio IRX (see Meurer et al. 1999; Calzetti 2000), that is also routinely exploited for high-redshift galaxies (see Bouwens et al. 2009, 2015, 2016a,b). For SFGs with intrinsic SFR $\dot M_\star\ga 30\, M_\odot$ yr$^{-1}$ the UV data, even when dust corrected via the UV slope-IRX relationship, strongly underestimate the intrinsic SFR, which is instead better probed by far-IR observations. This is because high SFRs occur primarily within heavily dust-enshrouded molecular clouds, while the UV slope mainly reflects the emission from stars obscured by the diffuse, cirrus dust component (Silva et al. 1998; Efstathiou et al. 2000; Efstathiou \& Rowan-Robinson 2003; Coppin et al. 2015; Reddy et al. 2015; Mancuso et al. 2016a). On the other hand, at low SFR $\dot M_\star\la 10\, M_\odot$ yr$^{-1}$ the dust-corrected UV data efficiently probe the intrinsic SFR. Moreover, in late-type galaxies at $z\la 1$ the far-IR emission itself can be contributed by the cirrus component, heated by the general radiation field from evolved stellar populations. To correct for such an effect, which otherwise may cause the SFR inferred from far-IR data to be appreciably overestimated, we have adopted the prescription by Clemens et al. (2013). In Fig.~\ref{fig|SFR_func} we report the overall data compilation from far-IR and dust-corrected UV observations. The luminosity $L$ and SFR $\dot M_\star$ scale have been related using $\log {\dot M_\star/ M_\odot~{\rm yr}^{-1}} \approx -9.8+\log {L/ L_\odot}$, a good approximation both for far-IR and (intrinsic) UV luminosities.

Then we have determined a smooth, analytic representation of the SFR function in terms of the standard Schechter shape
\begin{equation}
{{\rm d}N\over {\rm d}\log\dot M_\star}(\dot M_\star,z) = \mathcal{N}(z)\, \left[\dot M_\star\over\dot M_{\star, c}(z)\right]^{1-\alpha(z)}\,e^{-\dot M_\star/\dot M_{\star, c}(z)}~,
\end{equation}
characterized at any given redshift $z$ by three parameters, namely, the normalization $\mathcal{N}$, the characteristic SFR $\dot M_{\star, c}$ and the faint end slope $\alpha$. We determine the values of the three Schechter parameters over the range $z\sim 0-10$ in unitary redshift bins by performing an educated fit to the data. Specifically, UV data are fitted for SFRs $\dot M_\star\la 30\, M_{\odot}$ yr$^{-1}$ since in this range dust-corrections based on the $\beta_{\rm UV}$ are reliable, while far-IR data are fitted for SFRs $\dot M_\star\ga 10^2\, M_{\odot}$ yr$^{-1}$ since in this range dust emission is largely dominated by molecular clouds and reflects the ongoing SFR.  To obtain a smooth yet accurate representation of the SFR functions at any redshift, we find it necessary to (minimally) describe the redshift evolution for each parameter $p(z)$ of the Schechter function as third-order polynomial in log-redshift $p(z)=p_0+p_1\, \xi+p_2\,\xi^2+p_3\,\xi^3$, with $\xi=\log(1+z)$. The values of the parameters $\left\{p_i\right\}$ are reported in Table~1. The resulting SFR functions for representative redshifts $z\approx 0$, $1$, $3$, and $6$ are illustrated in Fig.~\ref{fig|SFR_func}.

In Mancuso et al. (2016a,b) and Lapi et al. (2017) we have validated the global SFR functions against independent datasets, including galaxy number counts at significative submm/far-IR wavelengths, redshift distributions of gravitationally lensed galaxies, galaxy stellar mass function via the continuity equation, main sequence of SFGs, cosmological evolution of the average SFR and GRB rates, and high-redshift observables including the history of cosmic reionization.

\subsection{SFR and BH accretion histories}\label{sec|timevo}

Second ingredient is constituted by deterministic evolutionary tracks for the history of star formation and BH accretion in an individual galaxy, gauged on a wealth of multi-wavelength observations and inspired theoretically by the in-situ coevolution scenario. This envisages star formation and BH accretion in galaxies to be essentially in situ, time-coordinated processes (e.g., Lapi et al. 2006, 2011, 2014, also Lilly et al. 2013; Aversa et al. 2015; Mancuso et al. 2016a,b), triggered by the early collapse of the host dark matter halos, but subsequently controlled by self-regulated baryonic physics and in particular by energy/momentum feedbacks from supernovae and AGNs.

In a nutshell, during the early stages of a galaxy's evolution, the competition between gas condensation and energy/momentum feedback from supernovae and stellar winds regulates the SFR. In low mass galaxies the SFR is small $\dot M_\star\la $a few tens $M_\odot$ yr$^{-1}$, and it slowly decreases over long timescales of several Gyrs because of progressive gas consumption. On the other hand, in high mass galaxies huge gas reservoirs can sustain violent, almost constant SFR $\dot M_\star\ga 10^2\, M_\odot$ yr$^{-1}$, while the ambient medium is quickly enriched with metals and dust; the galaxy behaves as a bright sub-mm/far-IR source. After a time $\tau_{\rm b}\sim$ some $10^8$ yr the SFR is abruptly quenched by the energy/momentum feedback from the central supermassive BH, and the environment is cleaned out; thereafter the stellar populations evolve passively and the galaxy becomes a ’red and dead’ early-type.

From the point of view of the central BH, during the early stages plenty of gas is available from the surroundings, so that considerable accretion rates sustain mildly super-Eddington emission with Eddington ratios $\lambda\equiv L/L_{\rm Edd}\ga 1$; radiation trapping and relativistic effects enforce  radiatively-inefficient, slim-disk conditions (see Begelman 1979; Li 2012; Madau et al. 2014). During these early stages the BH bolometric luminosity is substantially smaller than that of the host SFG, but increases exponentially. After a time $\tau_{\rm b}\ga $ a few $10^8$ yr, the nuclear power progressively increases to values similar or even exceeding that from star formation in the host galaxy. As mentioned above, strong energy/momentum feedback from the BH remove interstellar gas and dust while quenching star formation; the system behaves as an optical quasar. Residual gas present in the central regions of the galaxy can be accreted onto the BH at progressively lower, sub-Eddington accretion rates. When the Eddington ratio falls below a critical value around $\lambda\la 0.3$ (see McClintock et al. 2006) the disk becomes thin, yielding the standard SEDs observed in type-1 AGNs. Eventually, the BH activity ceases because of gas exhaustion in the nuclear region. At low-redshift $z\la 1$, especially within a rich environment, gravitational interaction or even a galaxy merger can temporarily re-kindle a starburst and the BH activity. A schematic evolution of the SFR and BH accretion rate as a function of the galaxy age is reported in Fig.~\ref{fig|timevo}.

On this basis, we have computed the relative time spent by the AGN in a given logarithmic bin of bolometric luminosity $L_{\rm AGN}$, i.e. the AGN duty cycle, as
\begin{equation}\label{eq|AGN_duty}
{{\rm d}\delta\over {\rm d}\log L_{\rm AGN}}(L_{\rm AGN},z|\dot M_\star)\approx {\tau_{\rm ef}+\tau_{\rm AGN}\over \tau_{\rm b}}\,\ln 10~;
\end{equation}
here $\tau_{\rm ef}$ is the $e-$folding time (depending on the Eddington ratio $\lambda$ and radiative efficiency) during the early AGN phase, $\tau_{\rm AGN}$ is the characteristic time of the declining AGN phase, and $\tau_{\rm b}$ is the duration of the star formation period before the AGN quenching. In Mancuso et al. (2016a,b) such parameters have been set by comparison with observations, including the Eddington ratio distributions at different redshift, the fraction of host SFGs in optically/X-ray selected quasars, the fraction of AGN hosts with given stellar mass as a function of the Eddington ratio, the BH mass function via the continuity equation, the main sequence of SFGs, and the AGN coevolution plane (i.e., bolometric AGN luminosity vs. SFR or stellar mass). Note that the AGN duty cycle depends on the average SFR through the above parameters, since at the end of the evolution of the galaxy the central BH to stellar mass ratio $M_{\rm BH}/M_\star$ must take on the locally observed values $\approx 10^{-3}$ (e.g., McConnell \& Ma 2013; Kormendy \& Ho 2013; Shankar et al. 2016). We defer the reader to the Mancuso et al. (2016b) paper for a detailed descriptions of the above parameter values and their dependence on the average SFR, that we adopt in full here.

\begin{figure*}
\epsscale{1}\plotone{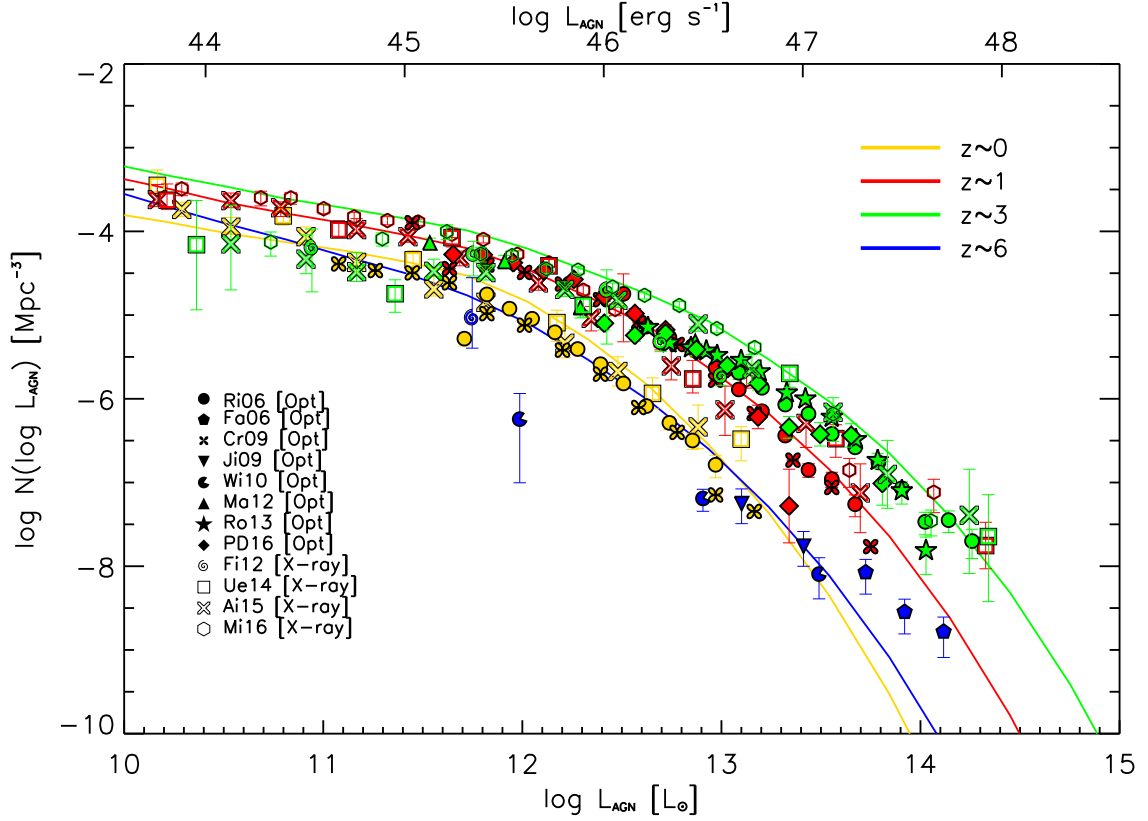}\caption{The (bolometric) AGN luminosity functions at redshifts $z=0$ (yellow), $1$ (red), $3$ (green) and $6$ (blue), as reconstructed from the SFR functions and the AGN duty cycle (see also Sect.~\ref{sec|basics}). Optical data (filled symbols) are from Richards et al. (2006; circles), Fan et al. (2006; pentagons), Croom et al. (2009; crosses), Jiang et al. (2009; inverse triangles), Willott et al. (2010; pacmans), Masters et al. (2012; triangles), Ross et al. (2013; stars), and Palanque-Delabrouille et al. (2016; diamonds); X-ray data (empty symbols) are from Fiore et al. (2012; spirals), Ueda et al. (2014; squares), Aird et al. (2015; big cross), and Miyaji et al. (2015; hexagons). The X-ray and optical luminosities of the data have been converted to bolometric by using the corrections from Hopkins et al. (2007), while the corresponding number densities have been corrected for obscured (including Compton thick) AGNs following Ueda et al. (2014).}\label{fig|AGN_LF}
\end{figure*}

In terms of the duty cycle, we can now map the SFR functions into the AGN bolometric luminosity functions as
\begin{equation}\label{eq|AGN_LF}
{{\rm d} N\over {\rm d}\log L_{\rm AGN}}(L_{\rm AGN},z) = \int{{\rm d}\log\dot M_\star}\,{{\rm d} N\over {\rm d}\log\dot M_\star}\,{{\rm d}\delta\over {\rm d}\log L_{\rm AGN}}(L_{\rm AGN},z|\dot M_\star)~.
\end{equation}
The outcome for representative redshifts $z\approx 0$, $1$, $3$, and $6$ is illustrated in Fig.~\ref{fig|AGN_LF}, and compared with a data compilation from optical and hard X-ray observations. The data have been converted to bolometric luminosity using the Hopkins et al. (2007) corrections, while the corresponding number densities have been corrected for obscured (also Compton thick) AGNs after Ueda et al. (2014). The pleasing agreement between our determination and the data confirms that the AGN duty cycle is correctly determined.

We anticipate that to properly address the radio emission in RQ systems, it will be convenient to compute the luminosity function of SFGs hosting an AGN with X-ray emission above a given threshold $L_{\rm X, min}$. This quantity is given by
\begin{equation}\label{eq|RS_SFR_func}
{{\rm d} N\over {\rm d}\log \dot M_\star}(\dot M_\star,z|>L_{\rm X,min}) = {{\rm d} N\over {\rm d}\log\dot M_\star}(\dot M_\star,z)\, \int_{>L_{\rm X,min}}{\rm d}\log L_{\rm AGN}\,{{\rm d}\delta\over {\rm d}\log L_{\rm AGN}}(L_{\rm AGN},z|\dot M_\star)~;
\end{equation}
the threshold $L_{\rm X, min}\approx 10^{42}$ erg s$^{-1}$ will be employed, since this is the value commonly adopted by observers (e.g., Padovani et al. 2015; Bonzini et al. 2013) to clearly discern the nuclear X-ray emission from that associated to star formation $L_{\rm X, SFR}\approx 7\times 10^{41}$ erg s$^{-1}\, (\dot M_\star/10^2\,M_\odot\, {\rm yr}^{-1})$, see e.g. Vattakunnel et al. (2012). The resulting SFR functions are illustrated as dashed lines in Fig.~\ref{fig|SFR_func}. Most of AGNs with significant X-ray powers are hosted at $z\ga 1$ by galaxies with SFRs $\dot M_\star\ga 10^2\, M_\odot$ yr$^{-1}$; however, their number density is smaller than that of the global star-forming population by a factor of $10^{-1}$, which reflects the nearly constant behavior of the SFR vs. the exponential growth of the BH accretion rate during most of the galaxy lifetime, quantified by the AGN duty cycle.

\section{Statistics of radio sources}\label{sec|radio}

In this section we discuss the radio emission properties of star-forming galaxies and radio AGNs, and compute the related contribution to the luminosity function, redshift distributions and counts at different radio frequencies.

\subsection{Star forming galaxies}

The radio emission associated to star formation comprises two components which are well known to correlate with the SFR (see Condon 1992; Murphy et al. 2011; Bressan et al. 2002): free-free emission (fully dominating at frequencies $\nu\ga 30$ GHz) emerging directly from HII regions containing massive, ionizing stars; synchrotron emission resulting from relativistic electrons accelerated by supernova remnants.

As to the free-free emission, we use the classic expression (see Murphy et al. 2011; Mancuso et al. 2015)
\begin{equation} \label{eq|Lff}
L_{\rm ff} \approx 3.75\times 10^{26}\, {\rm erg~s^{-1}~Hz^{-1}}\, {\dot M_\star\over M_\odot~{\rm yr}^{-1}}\, \left(T\over 10^4\, {\rm K}\right)^{0.3}\,g(\nu,T)\, e^{-h\nu/kT}
\end{equation}
where $g(\nu,T)$ is the Gaunt factor
\begin{equation}\label{eq|gaunt}
g(\nu,T)=\ln\left\{\exp\left[5.960-{\sqrt{3}\over \pi}\,\ln\left(Z_i\, {\nu\over {\rm GHz}}\,\left({T\over 10^4 K}\right)^{-1.5}\right)\right]+e\right\},
\end{equation}
approximated according to Draine (2011), and the quantity $e^{-h\nu/kT}$
tentatively renders electron energy losses. This equation reproduces the
Murphy et al. (2012) calibration at $33$ GHz for a pure hydrogen plasma ($Z_i=1$) with temperature $T\approx 10^4$ K; we adopt these values in the following.

As to the synchrotron emission, the calibration with the SFR is
a bit more controversial since it involves complex and poorly understood processes such as the production rate of relativistic electrons,
the fraction of them that can escape from the galaxy, and the magnetic
field strength. We use the calibration proposed by Murphy et al. (2011; 2012) and then adopted in the widely cited review by Kennicutt \& Evans (2012)
\begin{equation}\label{eq|Lsync}
L_{\rm sync}\approx 1.9 \times 10^{28}\, {\rm erg~s^{-1}~Hz^{-1}}\, {\dot M_\star\over M_\odot~{\rm yr}^{-1}}\, \left({\nu\over {\rm GHz}}\right)^{-\alpha_{\rm sync}}\, \left[1+\left({\nu\over 20\, {\rm GHz}}\right)^{0.5}\right]^{-1}\, F[\tau_{\rm sync}(\nu)]
\end{equation}
where $\alpha_{\rm sync}\approx 0.75$ is the spectral index (e.g., Condon 1992), the term in square brackets renders spectral ageing effects (see Banday \& Wolfendale 1991) and the function $F(x)=(1-e^{-x})/x$ takes into account synchrotron self-absorption in terms of the optical depth (e.g., Kellermann 1966; Tingay \& de Kool 2003)
\begin{equation}\label{eq|self}
\tau_{\rm sync}\approx (\nu/\nu_{\rm self})^{-\alpha_{\rm sync}-5/2}
\end{equation}
that is thought to become relevant at frequencies $\nu\la \nu_{\rm self}\approx 200$ MHz.

Given that for the Chabrier IMF the far-IR luminosity is given by $L_{\rm FIR}\,[{\rm W}]\approx 3\times 10^{36}\, \dot M_\star\,[M_\odot~{\rm yr}^{-1}]$ in terms of the SFR, the above calibrations Eqs.~(\ref{eq|Lff}) and (\ref{eq|Lsync}) yield a far-IR vs. $1.4$ GHz correlation parameter $q_{\rm FIR}\equiv \log(L_{\rm FIR}/3.75\times 10^{12}\, {\rm W})-\log(L_{\rm 1.4\, GHz}/{\rm W\, Hz^{-1}})\approx 2.77$; this is slightly higher than the classic value $q_{\rm FIR}\approx 2.35$ (see Yun et al. 2001) but in excellent agreement with the recent determinations by Novak et al. (2017) and Delhaize et al. (2017). Note that at $\nu\approx 1.4$ GHz a synchrotron to free-free luminosity ratio $L_{\rm synch}/L_{\rm ff}\approx 5.4$ is found, somewhat lower than the classic value $\approx 8$ quoted by Condon (1992) from his analysis of the M82 SED, but in good agreement with more recent data and models (see Murphy et al. 2011, 2012; Bressan et al. 2002; Obi et al. 2017).

Following Mancuso et al. (2015), we also take into account the lower efficiency in producing synchrotron emission by galaxies with small SFRs $\dot M_\star\la $ a few $M_\odot$ yr$^{-1}$ (cf. Bell 2003), on correcting Eq.~(\ref{eq|Lsync}) as
\begin{equation}\label{eq|Lsync_corr}
L_{\rm sync,corr}={L_{\rm sync}\over 1+ (L_{0,\rm sync}/L_{\rm synch})^{\zeta}}
\end{equation}
with $\zeta \approx 2$ and $L_{0,\rm sync}\approx 3\times 10^{28}$ erg s$^{-1}$ Hz$^{-1}$. We anticipate that this correction is necessary to reproduce the local $1.4$ GHz luminosity function at small radio powers $L_{1.4\, \rm GHz}\la L_{0,\rm sync}$.

Note that other phenomena may contribute to change the synchrotron luminosity in specific frequency and redshift range, that in the lack of a consensus physical understanding and detailed modeling we decide not to include in our fiducial approach. First, at low frequencies in a dense medium where relativistic and thermal electrons spatially coexist, the synchrotron emission can be absorbed; this would be described by an additional multiplicative factor $e^{-\tau_{\rm ff}}$ in Eq.~(\ref{eq|Lsync}), where
\begin{equation}\label{eq|tau_ff}
\tau_{\rm ff}\approx \left({T\over 10^4 K}\right)^{-1.35}\,\left({\nu\over 1.4\, {\rm GHz}}\right)^{-2.1}\, {{\rm EM}\over 6\times 10^6\, {\rm pc\, cm}^{-6}}
\end{equation}
is the free-free absorption optical depth (Condon 1992; Bressan et al. 2002) in terms of the emission measure EM of the plasma. However, on the average synchrotron emission in SFGs is produced on spatial scales much larger than that of the thermal electrons. Moreover, in most of the local galaxy population the physical conditions would imply free-free absorption to become relevant only at very low frequencies $\nu\la 100$ MHz, although there are some controversial cases related to starburst cores where absorption have been reported to be effective even at $\nu\la 1$ GHz (e.g., Vega et al. 2008; Schober et al. 2017). Note also that in high redshift SFGs the expected increase in the average density of the medium is easily offset by the $z-$dependence induced in the restframe $\tau_{\rm ff}\propto \nu^{-2.1}\,(1+z)^{-2.1}$.

Second, at high redshift the relativistic electrons producing synchrotron can lose energy due to a number of processes, most noticeably inverse Compton scattering off CMB photons (e.g., Murphy 2009; Lacki \& Thompson 2010; Schober et al. 2017) whose energy density grows with redshift as $(1+z)^4$. However, the amplitude of the breakdown and the redshifts at which it happens vary in connection with the assumed properties of high-redshift galaxies. In particular, Bonato et al. (2017) have shown that such effects are constrained to be minor at least out to redshift $z\la 3$. In fact, Smith et al. (2014) have measured a direct dependence of the radio to monochromatic far-IR luminosity ratio $L_{\rm 1.4\,GHz}/L_{\rm 250\,\mu m}$ with dust temperature, that could lead to some balancing of the inverse Compton losses. We stress that at high rest-frame frequencies the free-free emission (not affected by the above) will dominate anyway over the synchrotron, so that the correction to the total radio power will be small.

Third, an open issue concerns the evolution with redshift of the  normalization in the radio luminosity vs. SFR relation, namely the $q_{\rm FIR}$ parameter mentioned above. Some studies found it to be unchanged or to undergo only minor variations with redshift (e.g., Ibar et al. 2008; Bourne et al. 2011; Mao et al. 2011; Smith et al. 2014) while others have reported a significant, albeit weak, evolution (e.g. Seymour et al. 2009; Basu et al. 2015; Magnelli et al. 2015; Novak et al. 2017; Delhaize et al. 2017). This can be rendered on multiplying Eq.~(\ref{eq|Lsync}) by a factor $10^{q_0\, [1-(1+z)^{-q_1}]}$:  Magnelli et al. (2015) find $q_0=2.35$ and $q_1=0.12$, Novak et al. (2107) report $q_0=2.77$ and $q_1=0.14$, and Delhaize et al. (2017) suggest $q_0=2.9$ and $q_1=0.19$. On the other hand, Bonato et al. (2017) have shown that the evolution by Magnelli et al. (2015) is marginally consistent with the $1.4$ GHz radio luminosity functions and deep counts down to $\mu$Jy levels; in Sect.~\ref{sec|results} we will revise the issue in view of the most recent data on the redshift dependent radio luminosity function (Novak et al. 2017) and counts (Prandoni et al. 2017) of SFGs.

\begin{figure*}
\epsscale{0.8}\plotone{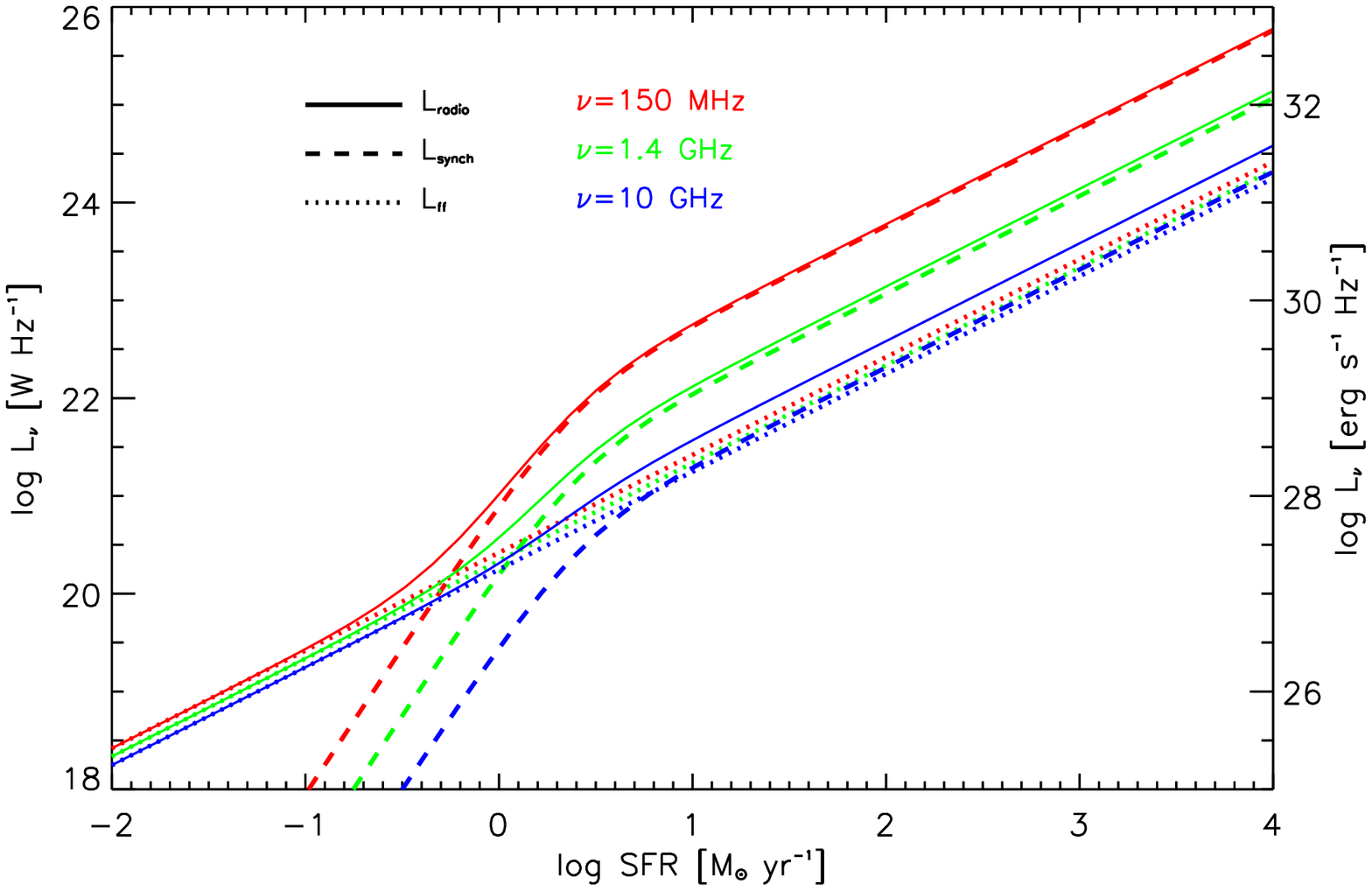}\\\plotone{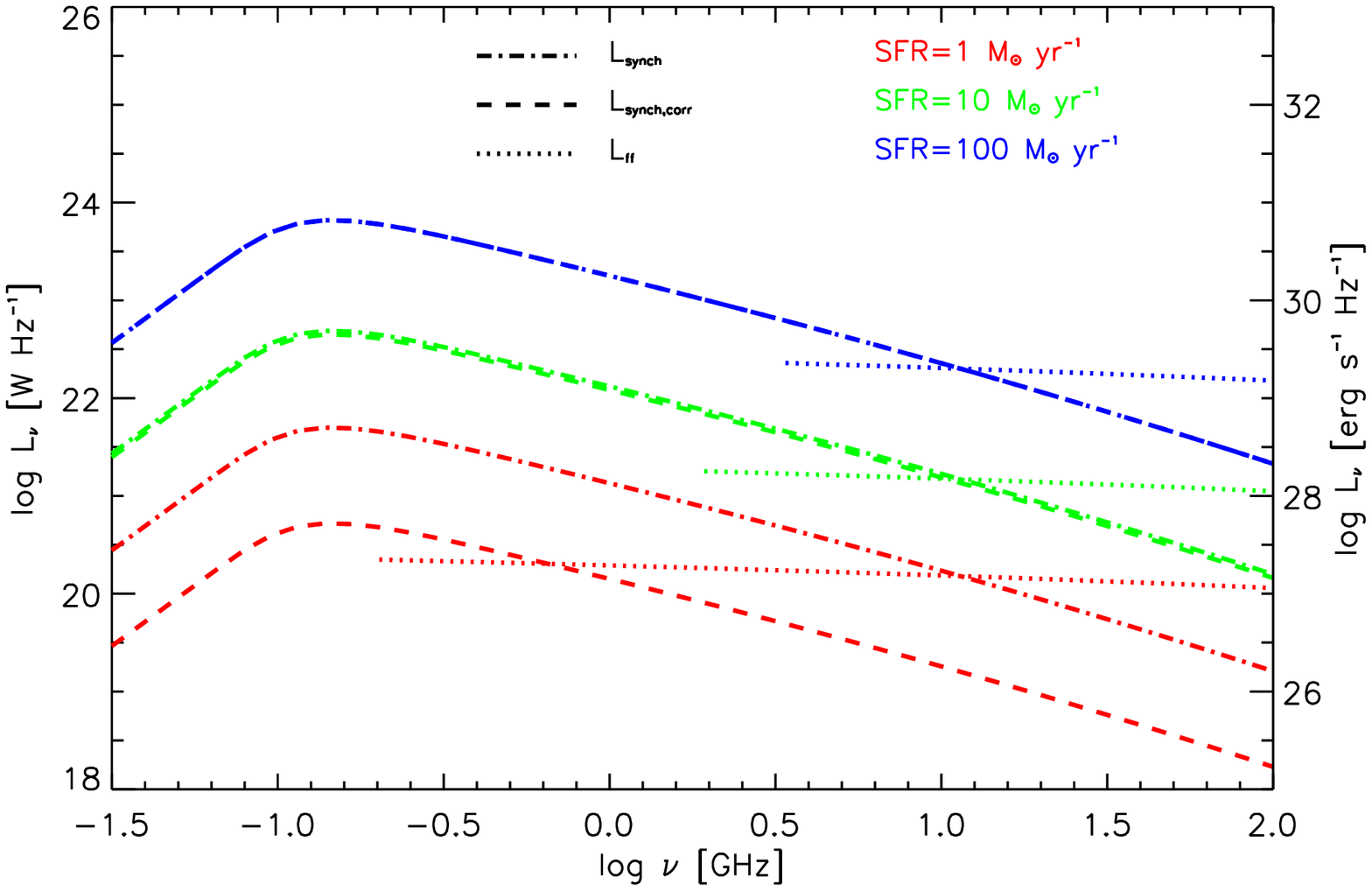}\caption{Top panel: synchrotron (dashed, including the suppression for small SFR after Eq.~\ref{eq|Lsync_corr}), free-free (dotted) and total radio (solid) luminosity from a SFG as a function of the SFR for three different frequencies $\nu\approx 0.15$ (red), $1.4$ (green), and $10$ GHz (blue). Bottom panel: synchrotron (dot-dashed lines refer to Eq.~\ref{eq|Lsync} while dashed lines includes the suppression for small SFR after Eq.~\ref{eq|Lsync_corr}; curves are distinguishable only for the smaller SFR) and free-free (dotted) luminosity as a function of frequency for three different values of the SFR $\dot M_\star\approx 1$ (red), $10$ (green), and $100\, M_\odot$ yr$^{-1}$ (blue).}\label{fig|Lradio}
\end{figure*}

The average total radio power for a given SFR is the sum of the contribution from synchrotron and free-free emission $\bar L_{\nu}(\dot M_\star) = L_{\rm synch,corr}+L_{\rm ff}$. In Fig.~\ref{fig|Lradio} we show these quantities as a function of the SFR for three representative frequencies $\nu\approx 0.15$, $1.4$, and $10$ GHz, and as a function of the frequency for three different values of the SFR. It is seen that the free-free emission increasingly dominates over synchrotron in mov ing toward high frequencies $\nu\ga 10$ GHz due to its flatter spectrum, and at small SFRs $\dot M_\star\la$ a few $M_\odot$ yr$^{-1}$ due to the inefficiency of synchrotron emission after Eq.~(\ref{eq|Lsync_corr}). At low frequencies $\nu\la 200$ MHz the synchrotron emission is also suppressed because of the self-absorption process.

We consider, consistently with observations (see Condon 1992; Bressan et al. 2002), a Gaussian scatter of $\sigma_{\log L}\approx 0.2$ dex around the average $\bar L_{\nu}(\dot M_\star)$ relationship. The radio luminosity function of SFGs is then obtained as
\begin{equation}
{{\rm d}N_{\rm SF}\over {\rm d}\log L_{\nu}}(L_\nu,z)={1\over \sqrt{2\pi}\, \sigma_{\log L}}\, \int{\rm d}\log \dot M_\star\, {{\rm d}N\over {\rm d}\log\dot M_\star}(\dot M_\star,z)\, e^{-[\log L_{\nu}-\log \bar L_{\nu}(\dot M_\star)]^2/2\,\sigma_{\log L}^2}
\end{equation}

\subsection{Radio-silent AGNs}\label{sec|RS}

According to the framework discussed in Sect.~\ref{sec|timevo} and illustrated in Fig.~\ref{fig|timevo}, during the early phase of a massive galaxy's evolution, the SFR is sustained at high, nearly constant values, while the BH mass is small but increases rapidly. After a few $e-$folding times, the X-ray power from the nucleus overwhelms that associated to star formation, so that the AGN is clearly detectable in X rays at luminosities $L_X\ga 10^{42}$ erg s$^{-1}$. This is the case for many SFGs selected in the far-IR band and then followed up in X-rays (e.g., Mullaney et al. 2012; Delvecchio et al. 2015; Rodighiero et al. 2015). However, the situation in the radio band is likely very different.

Since the BH accretes at high rates from large gas reservoirs, slim-disk conditions develop, featuring rather low radiative efficiency because of photon trapping and relativistic effects (see Begelman 1979; Li 2012; Madau et al. 2014); the accretion rates can be substantially super-Eddington $\dot M_{\rm BH}\gg L_{\rm Edd}/c^2$ but the emitted luminosity is only moderately above Eddington with $\lambda\equiv L/L_{\rm Edd}\ga$ a few. The accretion is nearly spherical and chaotic, hence the spin of the BH and of the disk stay small and rotational energy cannot be easily funnelled into jets to power radio emission via the Blandford \& Znajek (1977) or the Blandford \& Payne (1982) mechanisms (e.g., Meier 2002; Jester 2005; Fanidakis et al. 2011). In addition, in this phase the BH is growing but still too small to originate large-scale AGN outflows and winds (indeed star formation is ongoing in the host), so that even nuclear radio emission from AGN-driven shocks is not expected. In fact, absence of appreciable nuclear radio emission in strongly star-forming, gas-rich SFGs at high-redshift appears to be confirmed by recent observations (see Ma et al. 2016; Heywood et al. 2017).

All in all, in this phase the AGN may be detectable in X-rays but it is almost radio-silent, so that any radio emission from the system should be mainly ascribed to the SFR in the host SFG. The number density of SFGs hosting a radio-silent AGN detectable in X-rays can be easily estimated on the basis of Eq.~(\ref{eq|RS_SFR_func}) as
\begin{equation}\label{eq|RS_LF}
{{\rm d}N_{\rm RQ-SF}\over {\rm d}\log L_{\nu}}(L_\nu,z)={1\over \sqrt{2\pi}\, \sigma_{\log L}}\, \int{\rm d}\log \dot M_\star\, {{\rm d}N\over {\rm d}\log\dot M_\star}(\dot M_\star,z|>L_{X,\rm min})\, e^{-[\log L_{\nu}-\log \bar L_{\nu}(\dot M_\star)]^2/2\,\sigma_{\log L}^2}
\end{equation}
where $L_{X,\rm min}\approx 10^{42}$ erg s$^{-1}$ is usually considered by observers as the selection threshold for the presence of the AGN. We checked that adopting as a threshold the X-ray emission $L_{\rm X, SFR}\approx 7\times 10^{41}$ erg s$^{-1}\, (\dot M_\star/10^2\,M_\odot\, {\rm yr}^{-1})$ associated to star formation after the calibration by Vattakunnel et al. (2012) does not change appreciably the outcome.

\subsection{Radio-quiet AGNs}

During the late evolution of a massive galaxy with age exceeding some $10^8$ yr, the BH has grown to large masses, and originate outflows that can quench the star formation in the host. Meanwhile, the BH accretion rates decline to sub-Eddington levels, and the accretion disk becomes thin and radiatively efficient (see Shakura \& Sunyaev 1973). The BH and the accretion disk spin up rapidly and rotational energy can be easily funnelled into jets (e.g., Blandford \& Znajek 1977; Blandford \& Payne 1982).

\begin{figure*}
\epsscale{1}\plotone{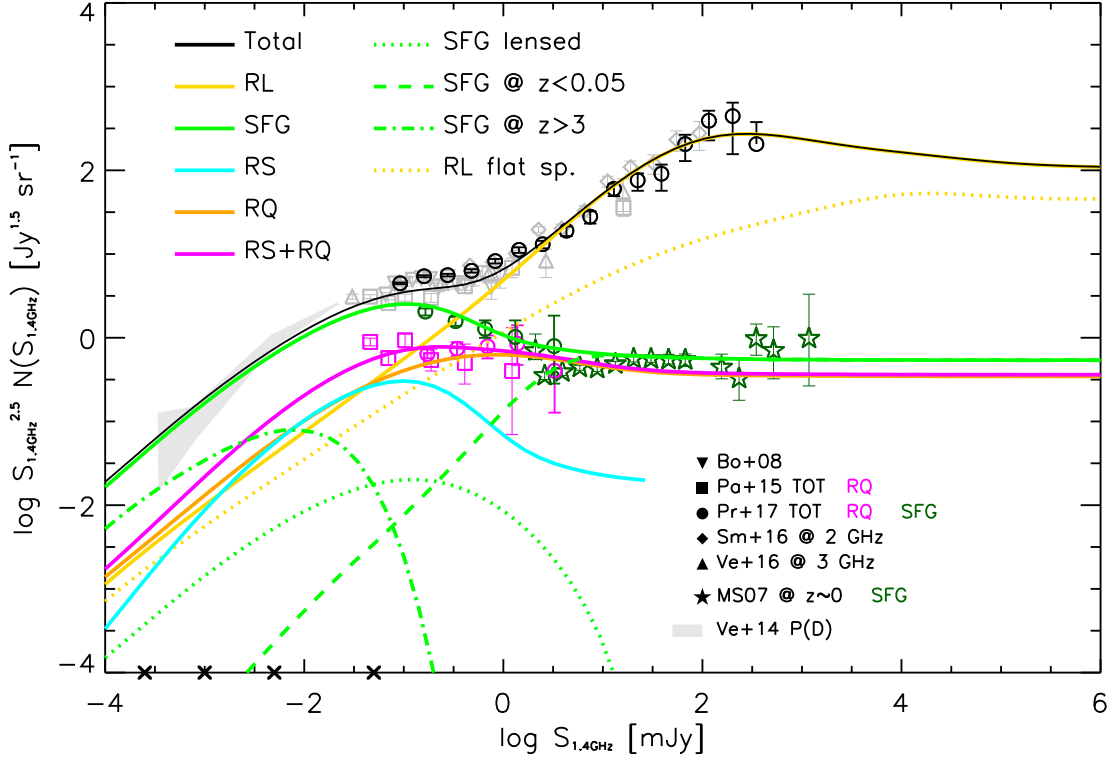}\caption{Euclidean number counts at $\nu= 1.4$ GHz, with contribution of different populations highlighted in color: green solid for SFG, cyan for RS AGNs, orange for RQ AGNs, magenta for RS+RQ AGNs, yellow for RL AGNs, and black for the total. The dotted, dashed and dot-dashed green lines illustrate the contribution of SFG that are strongly lensed, that are located at $z\la 0.05$ and that are at $z\ga 3$, respectively. The dotted yellow line highlights the negligible contribution from flat-spectrum RL AGNs (e.g., flat-spectrum radio quasars). Data are from Bondi et al. (2008, inverse triangles), Padovani et al. (2015, squares: grey for total and magenta for RQ AGNs), Prandoni et al. (2017, circles: grey for the total, green for SFGs and magenta for the RQ AGNs), Smolcic et al. (2016, diamonds, rescaled from $2$ GHz to $1.4$ GHz), Vernstrom et al. (2016, triangles, rescaled from $3$ GHz to $1.4$ GHz), Mauch \& Sadler (2007, stars: for SFGs at $z\sim 0$). The P(D) distribution from Vernstrom et al. (2014) is also shown as a grey shaded area at faint fluxes. The crosses on the abscissa indicate the flux limits for which the redshift distribution is shown in Fig.~\ref{fig|zdist_1.4GHz}.}\label{fig|counts_1.4GHz}
\end{figure*}

However, the jets driven by thin disk accretion are rather ineffective in producing radio emission with respect to the advection-dominated flows powering low-$z$, steep-spectrum RL objects\footnote{Generally speaking, the classic unification scheme for radio AGN comprises objects dominated by the core, beamed emission like Blazars (BL Lac and quasars at lower and higher power, respectively) and their counterparts observed at large viewing angles with respect to the jet axis (FRI and FRII radio galaxies at lower and higher powers, respectively). The radio power is also reasonably characterised by low and high excitation spectral lines (caveat the presence of obscuration). In the following we will also consider the distinction between flat and steep spectrum sources which, though being partly related to the observation frequency and selection criteria, is however needed in order to compute number counts statistics.} (e.g., Meier 2002; Jester 2005; Fanidakis et al. 2011). This appear to be confirmed by the observed anti-correlation between the ratio of the jet power to the accretion luminosity vs. the Eddington ratio (see see Punsly \& Zhang 2011; Fernandes et al. 2011; Sikora et al. 2013; Rusinek et al. 2017). In most of the instances, the net result of a thin disk accretion will be a RQ AGNs with weak, small-scale jets. Note, however, that when the BH mass is very large thin disk conditions may also originate a flat-spectrum RL AGN, as it can be the case for some blazars observed out to high redshift (see Ghisellini et al. 2013); on the other hand, these sources constitute a minority ($\la 10\%$) both of the radiatively-efficient AGNs and of the overall RL AGN population (cf. Sect.~\ref{sec|RL} and Fig.~\ref{fig|counts_1.4GHz}).

Alternatively to the small-scale jet origin, the radio emission of RQ AGNs may be traced back to shock fronts associated with the AGN-driven outflow (e.g., Zakamska \& Green 2014; Nims et al. 2015), or to winds originated from the outermost portion of the thin accretion disk (see Blundell et al. 2001; King et al. 2013), or to electron acceleration via magnetic reconnection in the thin disk corona (see Lahor \& Behar 2008; Raginski \& Laor 2016). Whatever mechanism operates, in this phase of a massive galaxy's evolution the radio emission from the system should be mainly ascribed to the nuclear activity typical of a RQ AGN.

The number density of RQ AGNs can be estimated as follows. We start from the AGN bolometric luminosity function of Eq.~(\ref{eq|AGN_LF}), convert the bolometric power in X-rays via the Hopkins et al. (2007) correction, and then derive the AGN radio power by using the relation between restframe X-ray and $1.4$ GHz radio luminosity observed for a sample of (mainly) RQ AGNs by Panessa et al. (2015; see also Brinkmann et al. 2000)
\begin{equation}
L_{\nu, \rm AGN}\approx 5.7\times 10^{-22}\, {\rm erg~s^{-1}~Hz^{-1}}\, \left({L_{X,\rm AGN}\over {\rm erg~s^{-1}}}\right)^{1.17}\,\left({\nu\over 1.4\, {\rm GHz}}\right)^{-\alpha_{\rm AGN}}
\end{equation}
where $\alpha_{\rm AGN}\approx 0.7$ is the spectral index for an optically thin synchrotron emission. Consistently with observations (e.g., Brinkmann et al. 2000; Panessa et al. 2015), we consider a scatter $\sigma_{\log L_X}\approx 0.4$ around the resulting average relationship $\bar L_{\nu, \rm AGN}(L_{\rm AGN})$. The related statistics is given by
\begin{eqnarray}
\nonumber {{\rm d}N_{\rm RQ-AGN}\over {\rm d}\log L_{\nu}}(L_\nu,z) &=&{1\over \sqrt{2\pi}\, \sigma_{\log L_X}}\, \int{\rm d}\log L_{\rm AGN}\, {{\rm d}N\over {\rm d}\log L_{\rm AGN}}(L_{\rm AGN},z)\times \\
\\
\nonumber &\times& e^{-[\log L_{\nu}-\log \bar L_{\nu,\rm AGN}(L_{\rm AGN})]^2/2\,\sigma_{\log L_X}^2}
\end{eqnarray}

\subsection{Radio-loud AGNs}\label{sec|RL}

Radio-loud AGNs are galaxies with clear signs of intense AGN activity in the radio band. In fact, RL systems constitute a mixed bag of objects with rather different properties in terms of accretion levels, excitation line emission, and variability timescales (see review by Tadhunter 2016; Padovani 2016).

Steep spectrum RL AGNs are characterized by optically thin synchrotron radio emission from (large-scale) relativistic jets; they are typically associated with low-redshift $z\la 1$ activity of very massive BHs at the center of massive early-type galaxies. These sources feature low Eddington ratios $\lambda\la 10^{-2}$ likely enforced by advection-dominated accretion flows (e.g., Narayan \& Yi 1994; Meier 2002; Fanidakis 2011). The associated large-scale jets have been shown to affect the thermodyamics of the surrounding intracluster medium (see reviews by McNamara \& Nulsen 2007 and Cavaliere \& Lapi 2013) and even to lift huge molecular gas reservoirs possibly promoting star formation (see Russell et al. 2017). Flat spectrum RL AGNs are instead characterized by more compact radio emission, typically associated to radiatively efficient, thin disk conditions (see review by Heckman \& Best 2014; also Massardi et al. 2016). 

RL AGNs are well known to dominate the bright portion of the radio counts above $0.5$ mJy at $1.4$ GHz. As such, they are not our main interest in this paper, which is mainly focused on the faint radio counts to be explored via next-generation surveys. However, for comparison with the total radio luminosity function and counts observed to date, we empirically include them in our analysis.

RL AGNs constitute a small fraction of the overall AGN population, reaching at most $\la 10\%$ for powerful optically or X-ray selected quasars (see Williams \& Rottgering 2015); for many of them the fueling mechanism is highly stochastic and orientation effects are relevant. Thus a description similar to that we pursued for RQ AGNs, based on the AGN bolometric luminosity function and on the radio vs. X-ray luminosity correlation (see Sambruna et al. 1999; Fan \& Bai 2016) is not viable. Therefore, we recur to the empirical description of the cosmological evolution for RL objects by Massardi et al. (2010), which has been extensively tested against a wealth of data on luminosity function and redshift distributions at least out to redshift $z\la 3$. For the reader's convenience we provide a brief account of the Massardi et al. (2010) description here.

These authors consider two flat-spectrum populations with different evolutionary properties, namely flat-spectrum radio quasars and BL Lacs, and a single steep-spectrum population; for sources of each population a simple power-law spectrum is adopted $S_\nu\propto \nu^{-\alpha}$ with $\alpha_{\rm FSRQ}=\alpha_{\rm BLLac}=0.1$ and $\alpha_{SS}=0.8$. The comoving luminosity function at a given redshift is described by a double-power law
\begin{equation}\label{eq|LF_RL}
{{\rm d}N\over {\rm d}\log L_\nu}(L_\nu,z) = {n_0\over [L_\nu/L_c(z)]^a+[L_\nu/L_c(z)]^b}
\end{equation}
and its redshift evolution is rendered in terms of a pure luminosity evolution
\begin{equation}
\log L_c(z) = \log L_c(0)+2\,k_{\rm ev}\,z\,z_{\rm top}\,[1-(z/z_{\rm top})^{m_{\rm ev}}/(1+m_{\rm ev})]
\end{equation}
where
\begin{equation}
z_{\rm top} = z_{\rm top,0} + {\delta z_{\rm top}\over 1+L_c(0)/L_\nu}
\end{equation}
is the redshift at which $L_c(z)$ reaches its maximum.

This empirical rendition is characterized by 8 parameters: $n_0$, $a$, $b$, $L_c(0)$, $k_{\rm ev}$, $m_{\rm ev}$, $z_{\rm top,0}$, $\delta z_{\rm top}$,
with different values for each of the three population considered (flat-spectrum radio quasars, BL Lac and steep spectrum). The parameters have been determined by Massardi et al. (2010) by fitting the luminosity function and redshift distributions from various surveys; we defer the reader to the Massardi et al. paper for a full description of this procedure, and for the resulting parameter values at $\nu=1.4$ GHz (see in particular their Table 1), that we adopt here.

It is worth mentioning that at frequencies $\nu\la 1$ GHz the RL counts turns out to be largely dominated by steep spectrum sources (see Fig.~\ref{fig|counts_1.4GHz}); for $\nu\ga$ a few GHz this is still true for $S_\nu\la 1$ Jy, while at higher fluxes flat spectrum sources starts to contribute appreciably.

\subsection{Number counts and redshift distributions}

We compute the differential number counts in the radio band by integrating over redshift the luminosity functions above
\begin{equation}\label{eq|counts}
{{\rm d}N\over {\rm d} \log S_{\nu}\,{\rm d}\Omega}(S_\nu)= \int{\rm d}z\, {{\rm d}V\over {\rm d}z\, {\rm d}\Omega}\, {{\rm d}N\over {\rm d}\log L_{\nu}}(L_{\nu\,(1+z)},z)
\end{equation}
where the flux is given by
\begin{equation}\label{eq|flux}
S_{\nu}= {L_{\nu\,(1+z)}\, (1+z)\over 4\pi\, D_L^2(z)}
\end{equation}
in terms of the cosmological volume per unit solid angle ${\rm d}V/{\rm d}z\, {\rm d}\Omega$ and of the luminosity distance $D_L(z)$. The redshift distribution is the integrand of the previous expression, in turn integrated over the luminosities above the one corresponding to a lower flux limit $S_{\nu,\rm lim}$ via Eq.~(\ref{eq|flux}).

We take into account strong galaxy-galaxy lensing of SFGs by using the amplification distribution ${\rm d}p/{\rm d}\mu$ derived in Lapi et al. (2012). The lensed differential counts are obtained as follows
\begin{equation}\label{eq|lenscounts}
{{\rm d}N_{\rm lens}\over {\rm d} \log S_{\nu}\,{\rm d}\Omega}(S_\nu) = \int{\rm d}z_s\,{1\over \langle\mu\rangle}\, \int^{\mu_{\rm max}}{\rm d}\mu\, {{\rm d}p\over {\rm d}\mu}\, {{\rm d}N\over {\rm d} \log S_{\nu}\,{\rm d}\Omega}(S_\nu/\mu)
\end{equation}
where the maximum amplification $\mu_{\rm max}\approx 25$ as appropriate for extended sources of a few kpcs is adopted; the factor $\langle\mu\rangle$ at the denominator can be approximated to $1$ in case of large area surveys, as considered here.

\section{Results}\label{sec|results}

\begin{figure*}
\epsscale{1}\plotone{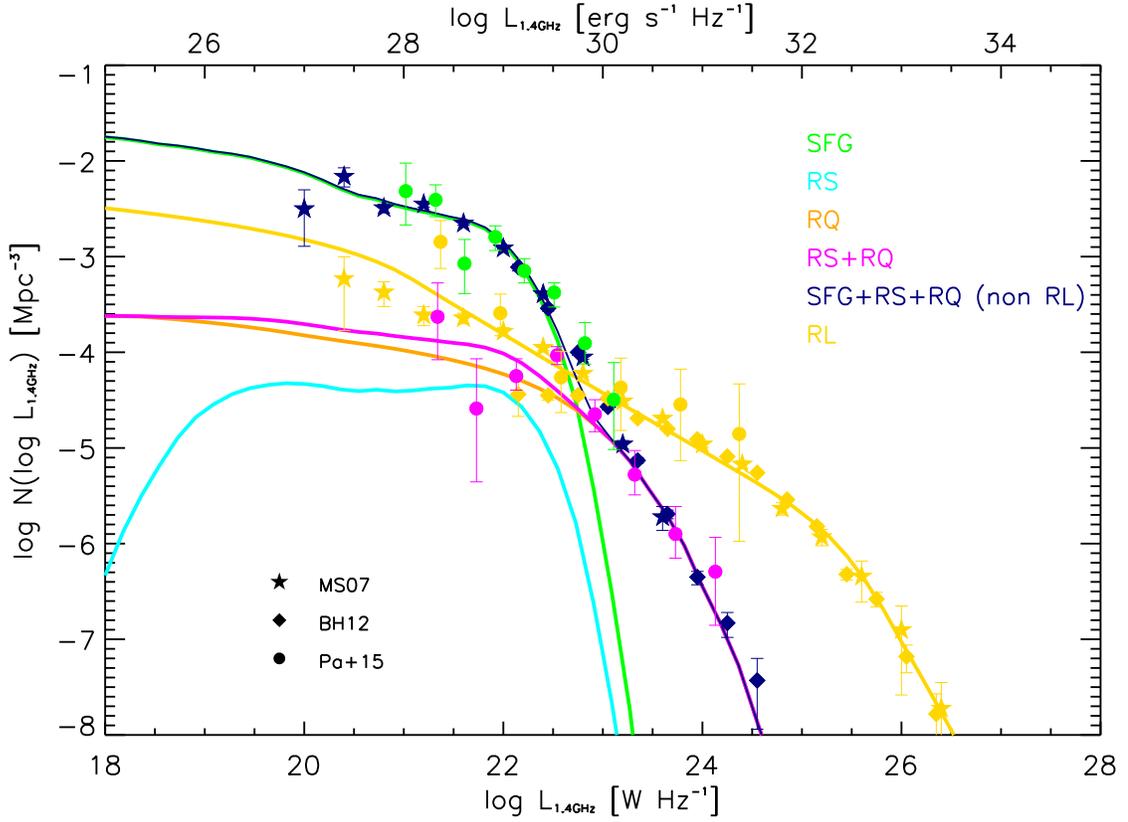}\caption{Local $z\sim 0$ radio luminosity function at $\nu=1.4$ GHz, with the contribution of different populations highlighted in color: green for SFGs, cyan for RS AGNs, orange for RQ AGNs, magenta for RS+RQ AGNs, navy for total of non RL sources, and yellow for RL AGNs. Data are from Mauch \& Sadler (2007, for RL and non-RL; stars), Best \& Heckman (2012, for RL and non-RL; diamonds), Padovani et al. (2015, for SFG, RQ and RL AGNs; circles).}\label{fig|LF_1.4GHz_z0}
\end{figure*}

In Fig.~\ref{fig|LF_1.4GHz_z0} we present the local radio luminosity function at $\nu=1.4$ GHz, with the contribution from the different populations of radio sources highlighted in color. SFGs account for the bulk of the local radio emission up to $L_{1.4\,\rm GHz}\la 10^{30}$ erg s$^{-1}$ Hz$^{-1}$. The fraction of such objects that contain an X-ray detectable (with threshold at $L_X \ga {\rm 10}^{42}$ erg s$^{-1}$) but radio-silent AGN is roughly $10^{-2}$ and marginally contributes to the radio luminosity function of RQ systems, which is instead dominated by nuclear emission from RQ AGNs. The latter dominates over the star-forming population for luminosities $L_{1.4\,\rm GHz}\ga 10^{30}$ erg s$^{-1}$ Hz$^{-1}$. Finally, RL systems  provides the bright tail of the radio luminosity function out to $L_{1.4\,\rm GHz}\la 10^{32}$ erg s$^{-1}$ Hz$^{-1}$. Our result agrees very well with the observational determination by Mauch \& Sadler (2007), Best \& Heckman (2012) and Padovani et al. (2015). The most relevant point here is that the luminosity function of non-RL sources is dominated by emission from star formation below, and nuclear emission from radio AGNs  above, the luminosity threshold of $L_{1.4\,\rm GHz}\sim 10^{30}$ erg s$^{-1}$ Hz$^{-1}$ (see also Kimball et al. 2011; Kellermann et al. 2016; White et al. 2017). For bright, but manifestly non-RL sources (discerned on the basis of, e.g.,  the $24\, \mu$m to $1.4$ GHz flux ratio), it will be important to test the presence of substantial radio-emission from the nucleus via future high-resolution observations.

\begin{figure*}
\epsscale{1.}\plotone{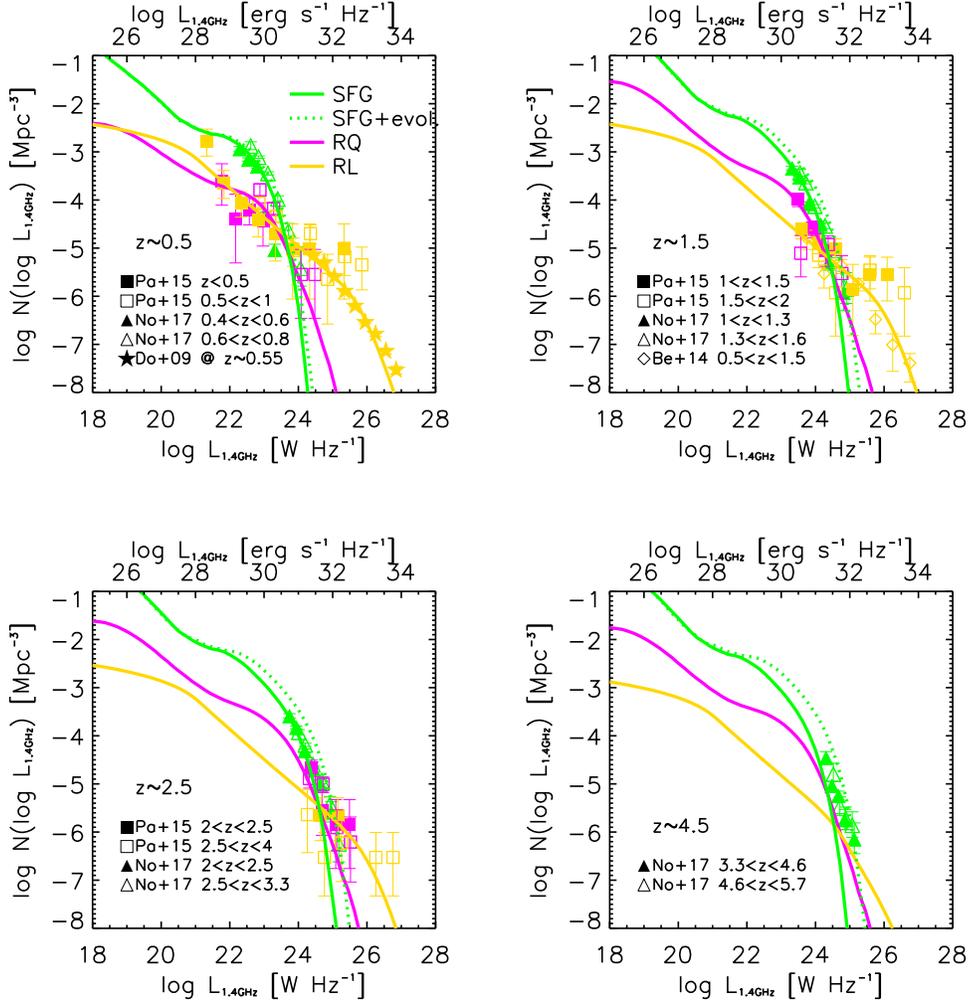}\caption{Radio Luminosity functions at $\nu= 1.4$ GHz for redshift $z\sim 0.5$ (top left panel), $1.5$ (top right), $2.5$ (bottom left) and $4.5$ (bottom right) with the contribution of different populations highlighted in color: green for SFGs, magenta for RQ AGNs and yellow for RL AGNs; the green dotted line includes the evolution in the normalization of the $L_{1.4\,\rm GHz}$ vs. SFR relation according to Novak et al. (2017; see also Delhaize et al. 2017). Data are from Padovani et al. (2015, squares), Novak et al. (2017, triangles), Donoso et al. (2009, stars), Best et al. (2014; diamonds).}\label{fig|LF_1.4GHz_RQ_z}
\end{figure*}

In Fig.~\ref{fig|LF_1.4GHz_RQ_z} we illustrate the $1.4$ GHz radio luminosity functions of the different populations at redshifts $z\approx 0.5$, $1.5$, $2.5$, and $4.5$. We compare our results with the observational determinations by Donoso et al. (2009), Best et al. (2014), Padovani et al. (2015) and Novak et al. (2017), finding a good agreement. Note that for $z\ga 1.5$ data on the faint end of the luminosity functions are still missing, so it will be crucial to obtain further observational constraints via the next generation ultra-deep radio surveys. The radio luminosity beyond which radio power is predominantly AGN-originated increases from $10^{30}$ erg s$^{-1}$ Hz$^{-1}$ at $z\approx 0$ to several $10^{31}$ erg s$^{-1}$ Hz$^{-1}$ at $z\approx 2.5$ and remain constant afterwards; this is mainly due to the strong cosmic evolution of the star formation rate function (cf. Fig.\ref{fig|SFR_func}; see also Gruppioni et al. 2015). This effect has been also pointed out in a different context by Magliocchetti et al. (2016).

In the same figure we also show the result for SFGs when including the evolution in the normalization of the $L_{1.4\, \rm GHz}$ vs. SFR relation as prescribed by Novak et al. (2017; see also Delhaize et al. 2017). The agreement with the observational data is appreciably improved only at $z\approx 4.5$ where, however, the evolution prescribed by Novak et al. is only extrapolated. All in all, we do not find clear evidence for a substantial evolution in the normalization of the $L_{\rm radio}$ vs. SFR relationship, though the data are still consistent with a weak evolution as claimed by Magnelli et al. (2015). By the same token, we do not find up to $z\la 4$ clear signs of an appreciable decrease in synchrotron luminosity due to energy losses of relativistic electrons via Compton upscattering off CMB photons.

In Fig.~\ref{fig|counts_1.4GHz} we show the Euclidean number counts at $\nu = 1.4$ GHz, with the contribution from different population highlighted as in the previous plots. As it is well known, the total counts for fluxes $S\ga 0.5$ mJy are dominated by low-redshift $z\la 1$ steep spectrum RL AGNs, with a minor contribution from flat-spectrum RL AGNs. At such bright fluxes, the non-RL sources are equally contributed by local SFGs and RQ AGNs; in particular, our counts for the low redshift $z\la 0.05$ SFG population are found to be in good agreement with the observational determination by Mauch \& Sadler (2007). At fluxes $S\la 0.5$ mJy, the total counts starts to be substantially contributed by evolving SFGs and RQ AGNs; the latter outnumber  RL sources below $0.1$ mJy. The agreement of our estimates for these population with the observational determinations by Prandoni et al. (2017; cf. also Padovani et al. 2015) is noticeable. In addition, the total counts (black) are in remarkable agreement with the observations by Prandoni et al. (2017), Smolcic et al. (2016) and Vernstrom et al. (2016). We stress that at around $S\sim 0.5$ mJy, a fraction of $1\%$ of the counts is contributed by strongly, lensed high-redshift SFGs, whose identification by followup observations will be extremely important. Finally, for $S\la 10^{-1}$ mJy SFGs dominates completely the total counts, that are found to be in good agreement with the preliminary determination based on the $P(D)$ distribution by Vernstrom et al. (2014). More than $30\%$ of the SFGs at these faint fluxes will be high-redshift $z\ga 3$ sources. Interestingly, we expect to still find a $10\%$ contribution to the total counts from RQ AGNs. High-redshift systems will be particularly interesting targets for the next generation of ultradeep radio continuum surveys to be conducted with SKA and its precursors (see Mancuso et al. 2015).

\begin{figure*}
\epsscale{1}\plotone{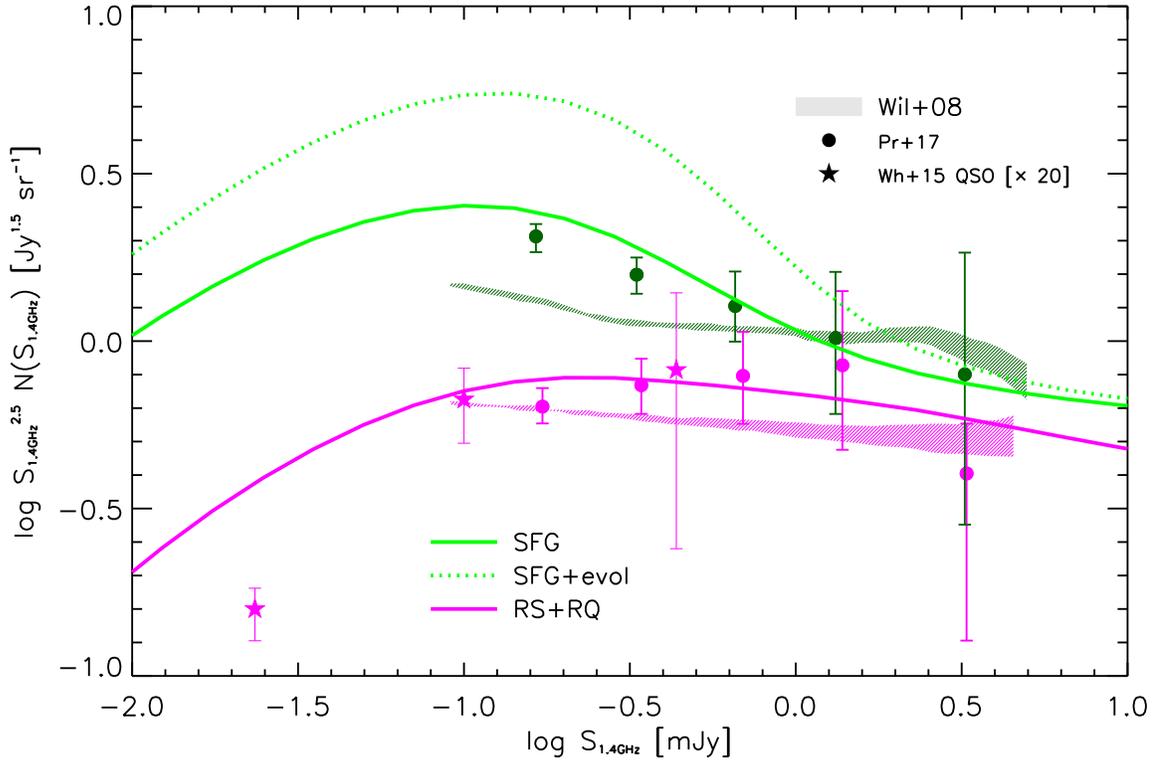}\caption{Euclidean number counts at $\nu= 1.4$ GHz contributed by SFGs (green) and RS+RQ AGNs (magenta). The green dotted line includes the evolution in the $L_{1.4\, \rm GHz}$ vs. SFR relation as prescribed by Novak et al. (2017; see also Delhaize et al. 2017). Shaded areas are from the S3-SEX semiempirical simulations by Wilman et al. (2008), and take into account cosmic variance effects on a $5$ deg$^2$ field. Data are from Prandoni et al. (2017; circles), and from White et al. (2015; stars; rescaled upwards by a factor $20$ to highlight the shape).}\label{fig|counts_1.4GHz_comp}
\end{figure*}

In Fig.~\ref{fig|counts_1.4GHz_comp} we zoom on our results for the counts of SFGs and RQ AGNs, and compare them in detail to the recent LH data by Prandoni et al. (2017) and to the outcome from the semi-empirical sky simulation developed in the framework of the SKA Simulated Skies project (S3-SEX; see Wilman et al. 2008). All in all, our results pleasingly agree with the data both for SFGs and for RQ AGNs, especially when considering that the lowest flux bins of the Prandoni et al. (2017) counts are affected by some incompleteness. Our approach performs comparably to the current S3-SEX model on RQ AGNs and significantly better on SFGs.

We also show the result for SFGs when including the evolution in the normalization of the $L_{1.4\, \rm GHz}$ vs. SFR relation as prescribed by Novak et al. (2017; see also Delhaize et al. 2017). The agreement with the observational data by Prandoni et al. (2017) for SFG is substantially worsened, although the total counts (considering the contribution of RL AGNs) are still consistent with the data. As suggested by Delhaize et al. (2017), this could be indication that AGN contribution may bias the evolutionary trend observed in the $L_{1.4\, \rm GHz}$ vs. SFR relationship.

Moreover, in the figure we report the data by White et al. (2015; see also White et al. 2017). Basing on a sample of mostly unobscured RQ quasars at $z\la 3$ from the VIDEO survey, these authors claimed that the shape of their counts is suggestive of a nuclear origin for the radio emission of these objects. As suggested by White et al. (2015) it is difficult to impose in theoretical approaches the same criteria used for their quasar selection, so the focus should be on the shape of their results more than on the normalization. Thus we rescale upwards their data by a factor $20$ to  highlight that their shape is similar to our result for RQ AGNs, but differs substantially from that of SFGs. This finding is consistent with our scenario for radio emission from SFGs and AGNs as discussed in Sect.~\ref{sec|radio} and illustrated in Fig.~\ref{fig|timevo}. An appreciable fraction ($\ga 70\%$, see Omont et al. 2003; Netzer et al. 2016; Harris et al. 2016) of optically-selected quasars is constituted by objects caught after the AGN luminosity peak, when the SFR may be decreased by AGN feedback while conditions of thin disk accretion onto the BH, conducive to nuclear radio emission, have set in.

Note, however, that the conclusions by White et al. (2015, 2017) regarding the nuclear origin in the radio emission of RQ systems are somewhat driven by their selection criteria, that tend to pick up radio powers larger than several $10^{30}$ erg s$^{-1}$ Hz$^{-1}$, a portion of the radio luminosity function populated by RQ AGNs (cf. Fig.~\ref{fig|LF_1.4GHz_z0}). We expect that a fraction around $30\%$ of optically selected quasars is constituted by objects caught before or soon after the AGN luminosity peak; as such they feature still sustained star formation activity, that can dominate radio emission at levels $\la$ a few $10^{30}$ erg s$^{-1}$ Hz$^{-1}$, as found by Kimball et al. (2011) and Kellermann et al. (2016).

\begin{figure*}
\epsscale{1}\plotone{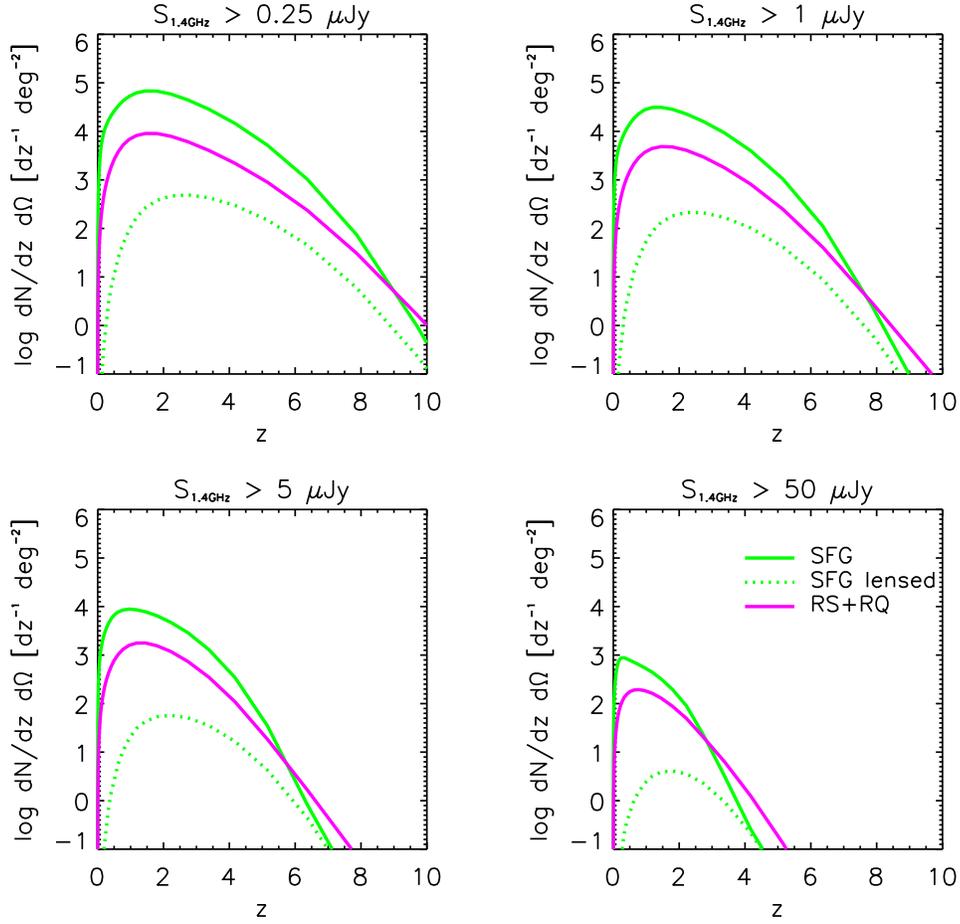}\caption{Redshift distributions at $\nu = 1.4$ GHz, with the contribution from different populations highlighted in colors: green for SFGs (solid line for overall and dotted line for lensed population), and magenta for RS+RQ AGNs. The panels refer to $4$ different flux limits $S_{1.4\, \rm GHz}\ga 0.25$, $1$, $5$, and $50\, \mu$Jy representative of surveys to be conducted by the SKA1-MID and its precursors.}\label{fig|zdist_1.4GHz}
\end{figure*}

In Fig.~\ref{fig|zdist_1.4GHz} we present the redshift distributions at $1.4$ GHz, for different flux limits $S_{1.4\, \rm GHz} \ga 0.25\, \mu$Jy, $1\,\mu$Jy, $5\,\mu$Jy and $50\,\mu$Jy representative of surveys planned on SKA1-MID (wide, deep and ultra deep surveys; see Prandoni \& Seymour 2015) and its precursors like EMU on ASKAP or MIGHTEE on MeerKat (see Norris et al. 2013 for an overview of ongoing or planned surveys with SKA pathfinders and precursors). As expected the bulk of the distributions at these faint fluxes is provided by SFGs, with an increasing contributions of RQ AGNs at high-redshift. The fraction of strongly lensed SFGs increases from $1\%$ at $z\approx 2$ to $10\%$ at $z\ga 6$.

\begin{figure*}
\epsscale{1}\plotone{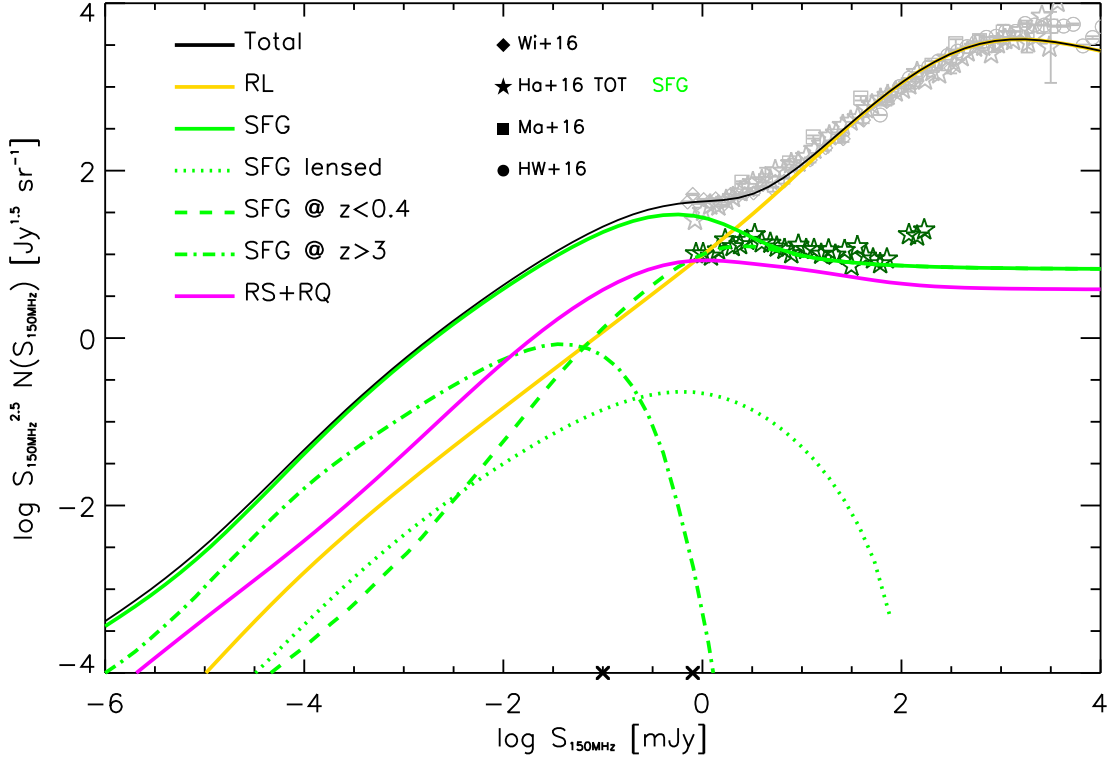}\caption{Euclidean number counts of different populations at $\nu= 150$ MHz, with contribution of different populations highlighted in color: green solid for SFGs, magenta for RS+RQ AGNs, yellow for RL AGNs, and black for the total. The dotted, dashed and dot-dashed green lines illustrate the contribution of SFGs that are gravitationally lensed, that are located at $z\la 0.4$ and that are at $z\ga 3$, respectively. Data are from Williams et al. (2016, diamonds), Hardcastle et al. (2016, stars: grey for the total and green for local SFGs), Mahony et al. (2016, squares), and Hurley-Walker et al. (2016, circles). The crosses on the abscissa indicate the flux limits for which the redshift distribution is shown in Fig.~\ref{fig|zdist_150MHz}.}\label{fig|counts_150MHz}
\end{figure*}

In Fig.~\ref{fig|counts_150MHz} we plot the Euclidean normalized number counts at $\nu=150$ MHz, the baseline working frequency of LOFAR. Remarkably, we find that at such low frequencies the synchrotron self-absorption plays an important role in shaping the normalization of the euclidean part of the counts, that is contributed from low-redshift SFGs. Specifically, by comparing with the observational determination for SFGs at $z\la 0.4$ by Hardcastle et al. (2016), we determine the average value $\nu_{\rm self}\approx 120\pm 50$ MHz for the characteristic frequency of the self-synchrotron emission appearing in Eq.~(\ref{eq|self}). Our estimate of the total counts agree very well with the data from Williams et al. (2016), Hardcastle et al. (2016), Mahony et al. (2016) and Hurley-Walker et al. (2017; see also Franzen et al. 2016). We stress that present data probe the counts down to $S_{150\, \rm MHz}\ga 1$ mJy, where they are mostly contributed by low-redshift $z\la 1$, steep-spectrum RL AGNs. However, the planned deep tiers of the LOFAR survey (see Rottgering 2010), and even more SKA surveys (Prandoni \& Seymour 2015), are expected to improve the flux limit by a factor larger than $10$, where the contribution from high-redshift SFGs and RQ AGNs will take over.

\begin{figure*}
\epsscale{1}\plotone{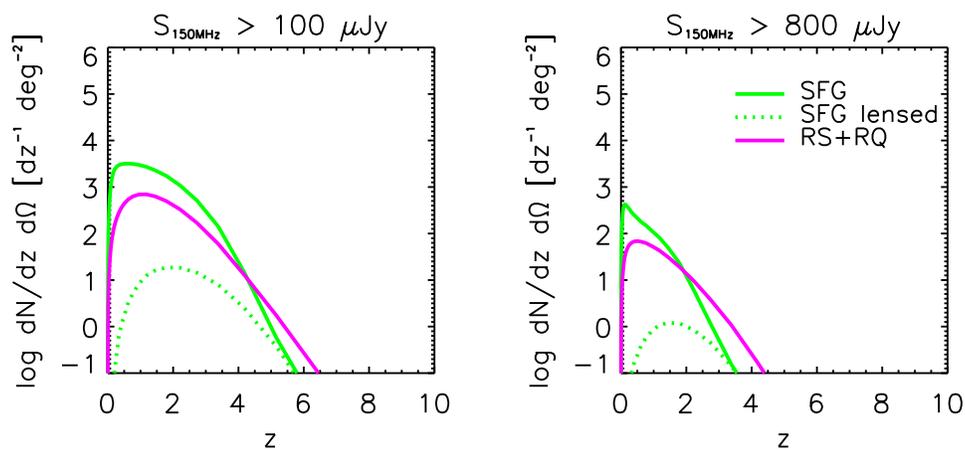}\caption{Redshift distributions at $\nu=150$ MHz, with the contribution from different populations highlighted in colors: green for SFGs (solid line for overall and dotted line for lensed population), and magenta for RS+RQ AGNs. Different panels refer to $2$ different flux limits $S_{150\, \rm MHz}\ga 100$ and $800\, \mu$Jy representative of surveys conducted by LOFAR or planned on LOFAR and SKA.}\label{fig|zdist_150MHz}
\end{figure*}

The situation is clearer in Fig.~\ref{fig|zdist_150MHz}, that shows the redshift distributions at $\nu=150$ MHz at the current level around $S_{150\,\rm MHz}\ga 800\, \mu$Jy and the prospective one at $100\,\mu$Jy. At the fainter flux limit, the contribution of RQ AGNs will start to dominate over the SFG population at $z\ga 4$; thus these low-frequency observations, when performed over wide areas like in future SKA surveys, can be effectively used to look for RQ AGNs at very high redshift, even out to the epoch of cosmic reionization.

\begin{figure*}
\epsscale{1}\plotone{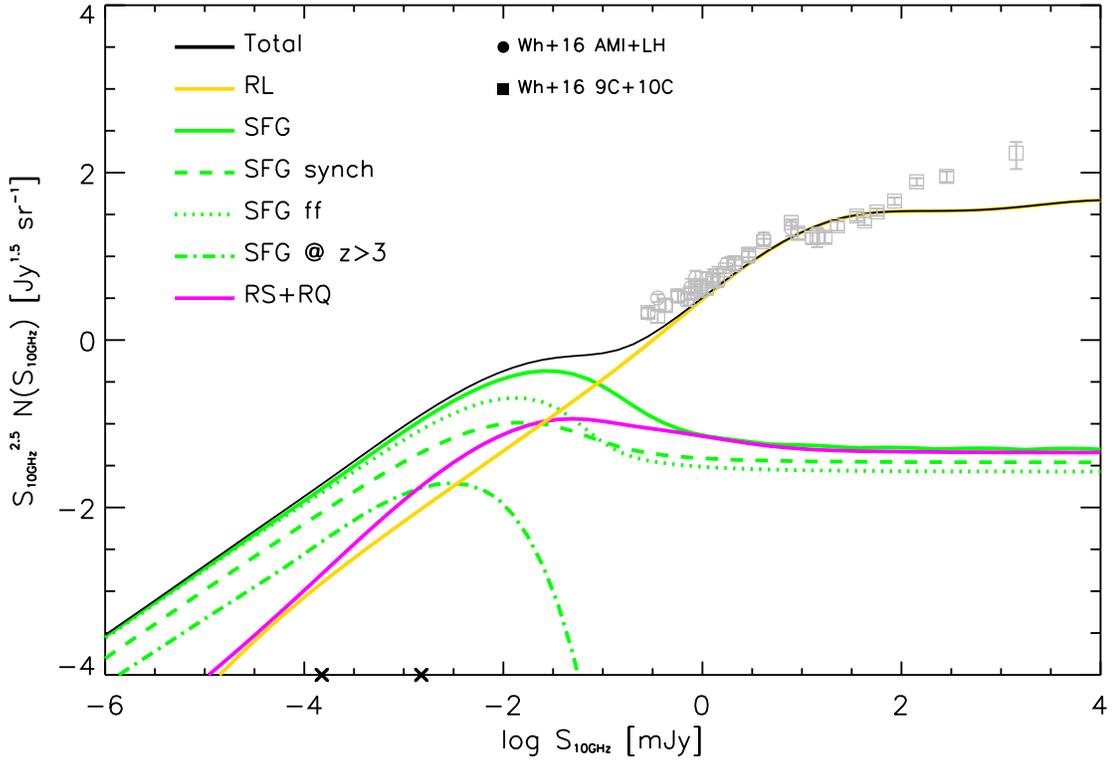}\caption{Euclidean number counts at $\nu=10$ GHz, with contribution of different populations highlighted in color: green solid for SFGs, magenta for RS+RQ AGNs, yellow for RL AGNs, and black for the total. The dashed and dotted green lines illustrate the average contribution from synchrotron and free-free to the emission of the SFG population. The dot-dashed green line refers to SFGs located at $z\ga 3$. Data are from Whittam et al. (2016, circles for AMI$+$LH and squares for 9C$+$10C fields, rescaled from $15.7$ to $10$ GHz). The crosses indicate the flux limits for which the redshift distribution is shown in Fig.~\ref{fig|zdist_10GHz}.}\label{fig|counts_10GHz}
\end{figure*}

\begin{figure*}
\epsscale{1}\plotone{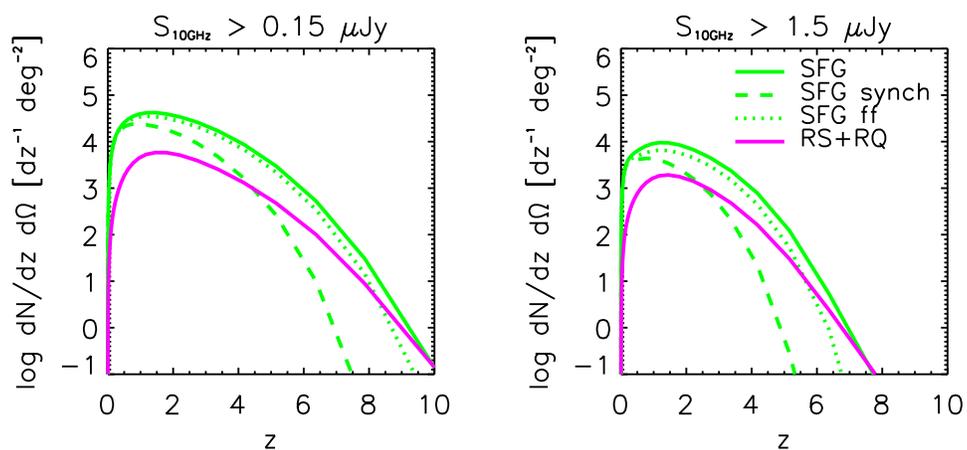}\caption{Redshift distributions at $\nu= 10$ GHz, with the contribution from different populations highlighted in colors: green for SFG (solid line for overall, dotted for free-free and dashed line for synchrotron emission), magenta for RS+RQ AGNs. Different panels refer to $2$ different flux limits $S_{15\, \rm GHz}\ga 0.15$ and $1.5\, \mu$Jy representative of surveys to be conducted by the SKA.}\label{fig|zdist_10GHz}
\end{figure*}

In Fig.~\ref{fig|counts_10GHz} we show the Euclidean number counts at $\nu = 10$ GHz, a high frequency that will be covered by the future SKA1-MID surveys. Our predictions for the total counts well agrees with the data from Whittam et al. (2016), that comprise the AMI, LH, 9C and 10C fields (rescaled from $15.7$ to $10$ GHz). The reader may appreciate that current data are dominated by RL sources. However, future SKA surveys can probe the counts down to flux limits of $S_{\rm 15 GHz} \ga 0.15$ and $1.5\, \mu$Jy (Prandoni \& Seymour 2015), where the contribution of high-redshift SFGs and RQ AGNs will take over; the corresponding redshift distributions are plotted in Fig.~\ref{fig|zdist_10GHz}. We highlight that the emission from SFGs at $10$ GHz for bright fluxes is almost equally contributed by synchrotron and free-free (see also Bressan et al. 2002; Obi et al. 2017), with the latter becoming increasingly dominant in the sub-mJy range; in terms of redshift, the free-free emission starts to take over for $z\ga 1-2$. Note that $\nu\approx 10$ GHz is an optimal frequency to probe the free-free emission over an extended redshift range $z\sim 1-8$ since there it dominates over synchrotron, and it is redshifted to restframe frequencies appreciably below $\nu\approx 150$ GHz where dust emission starts to become relevant. Moreover, given that the synchrotron emission at high redshift $z\ga 6$ may be considerably affected by inverse Compton scattering off the CMB photons, high frequency surveys at $\nu\ga 10$ GHz can be extremely useful to search for high-$z$ SFGs via their free-free emission.

\subsection{Further observational constraints}\label{sec|tests}

It will be of fundamental importance to test our expectations of Sect.~\ref{sec|radio} concerning the radio emission from SFGs and radio AGNs in different stages of a massive galaxy's evolution, by looking at large samples of radio sources with multi-band coverage (e.g., X-ray, far-IR and radio). We present below three specific examples.

\begin{figure*}
\epsscale{0.8}\plotone{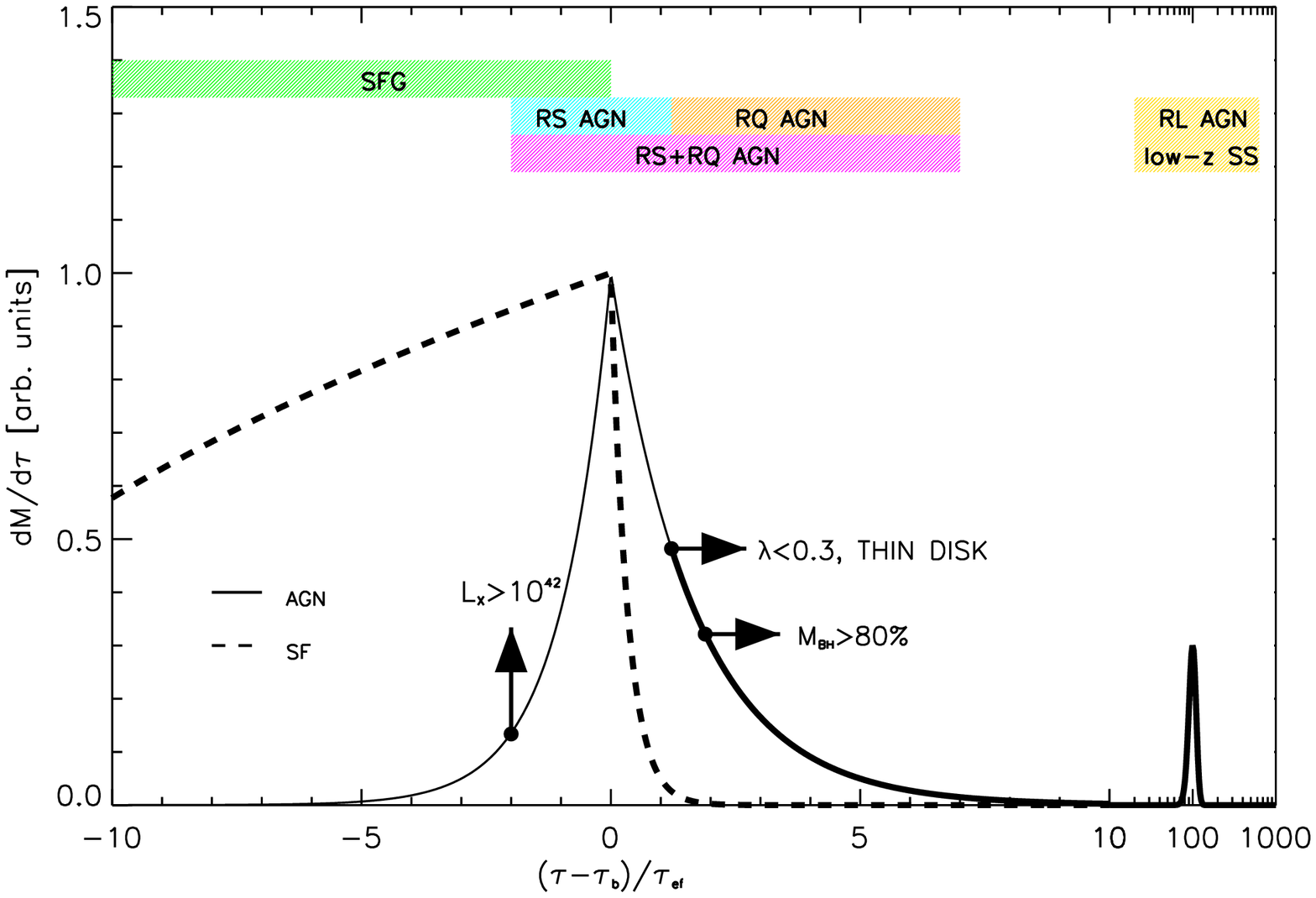}\plotone{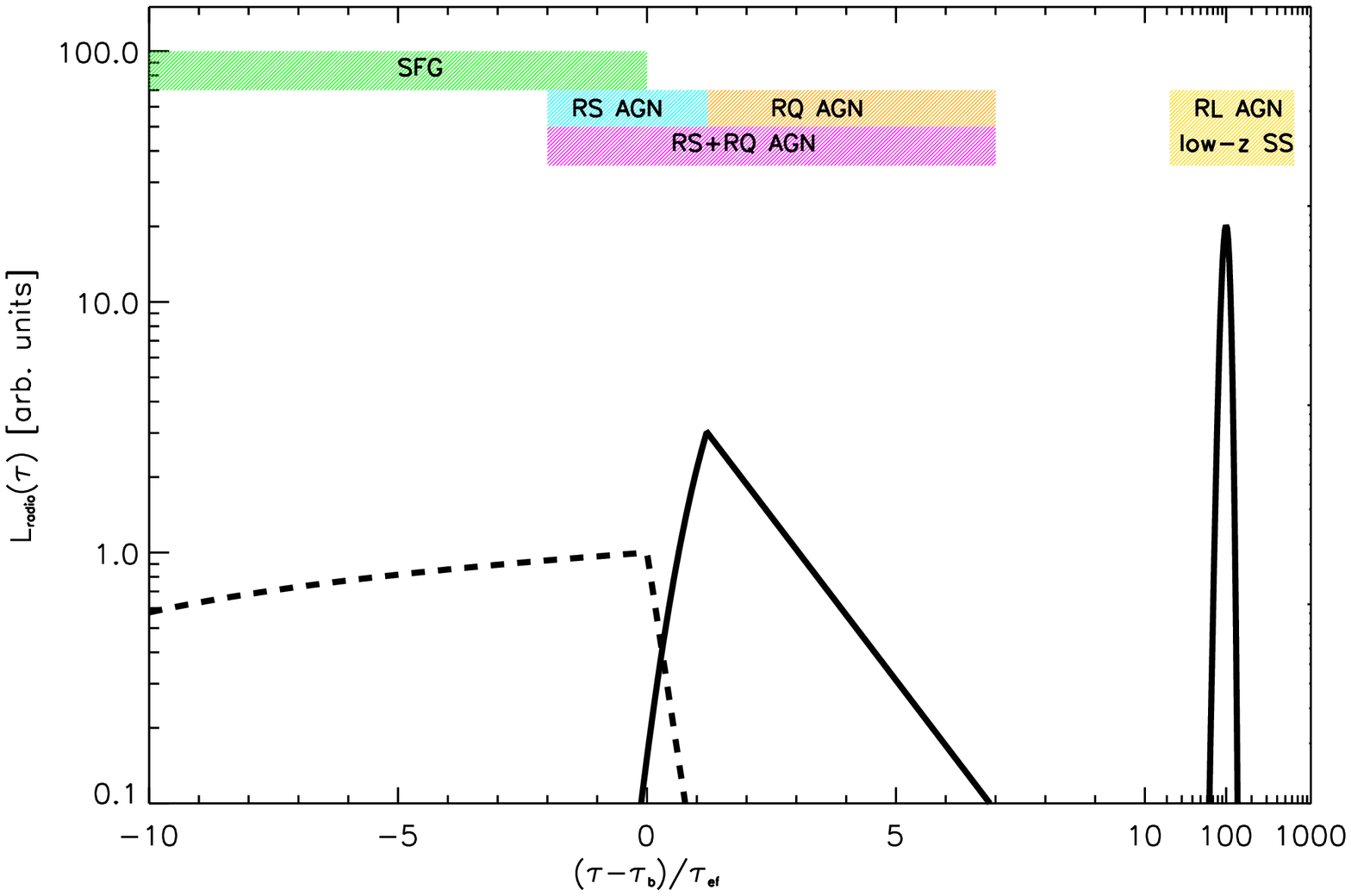}\caption{Top panel: schematic evolution with galactic age (in units of the BH $e-$folding time, amounting to some $10^7$ yr) of the SFR (dashed) and of the BH accretion rate (solid). The BH curve is thin where the AGN is radio-silent, and thick where it is radio-active. The dots with arrows indicate the epochs when: (i) the X-ray AGN luminosity exceeds $10^{42}$ erg s$^{-1}$, so that nuclear activity is detectable; (ii) the transition from a radio-silent slim-disk accretion to a radio-quiet, thin-disk accretion at $\lambda\la 0.3$ sets in; (iii) 80\% of the BH mass has been accumulated. Bottom panel: corresponding evolution of the radio luminosity associated to star formation and AGN emission. In both panels the colored strips indicate the different evolutionary stages in terms of the radio emission from the system: green refer to radio-emitting star-forming galaxies (SFGs), cyan to radio-silent (RS) AGNs, orange to radio-quiet (RQ) AGNs, magenta to RS+RQ AGNs, and yellow to low-$z\la 1$, steep spectrum radio-loud (RL) AGNs associated with a late-time activity at low accretion rates.}\label{fig|timevo}
\end{figure*}

\begin{figure*}
\epsscale{0.8}\plotone{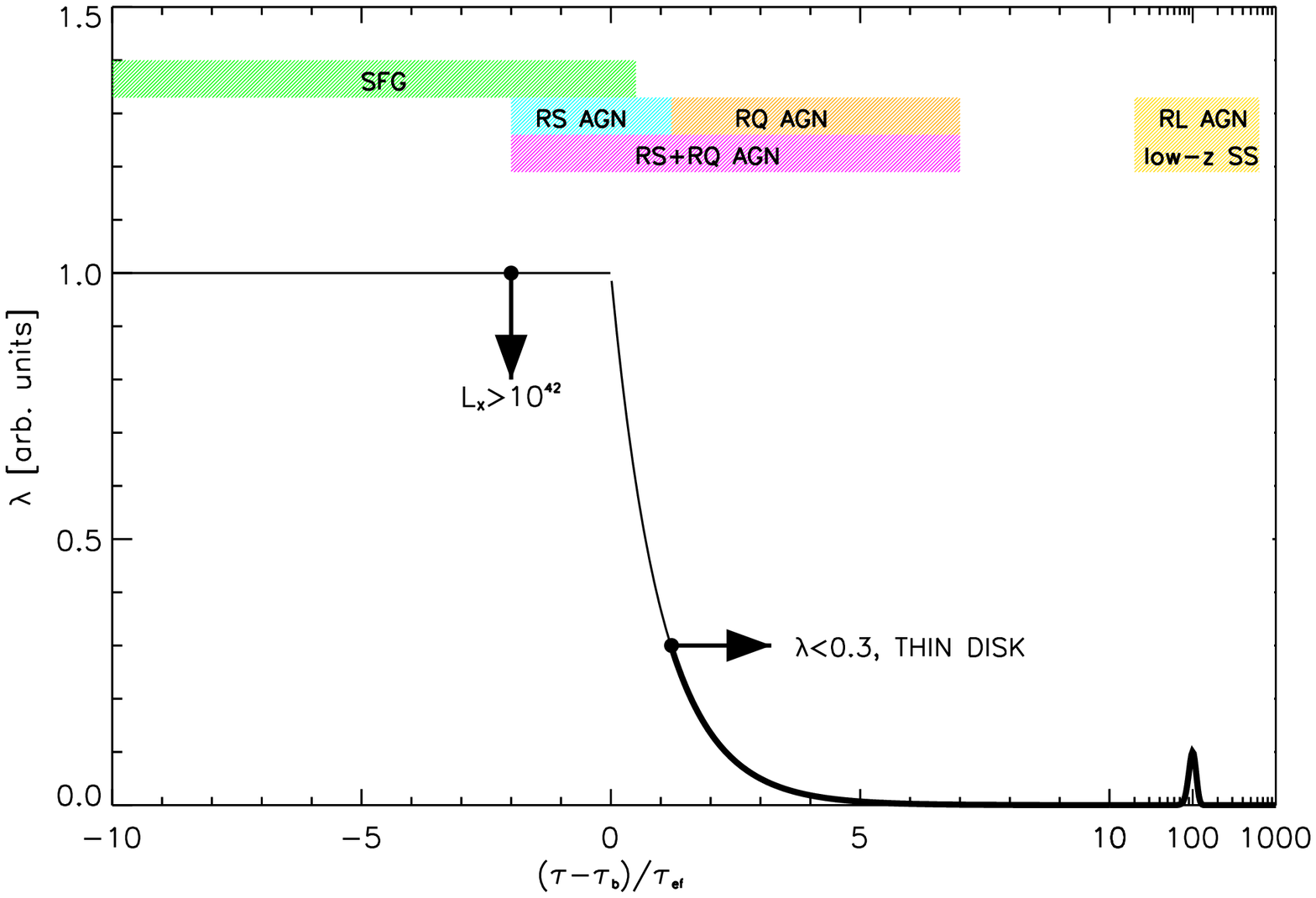}\plotone{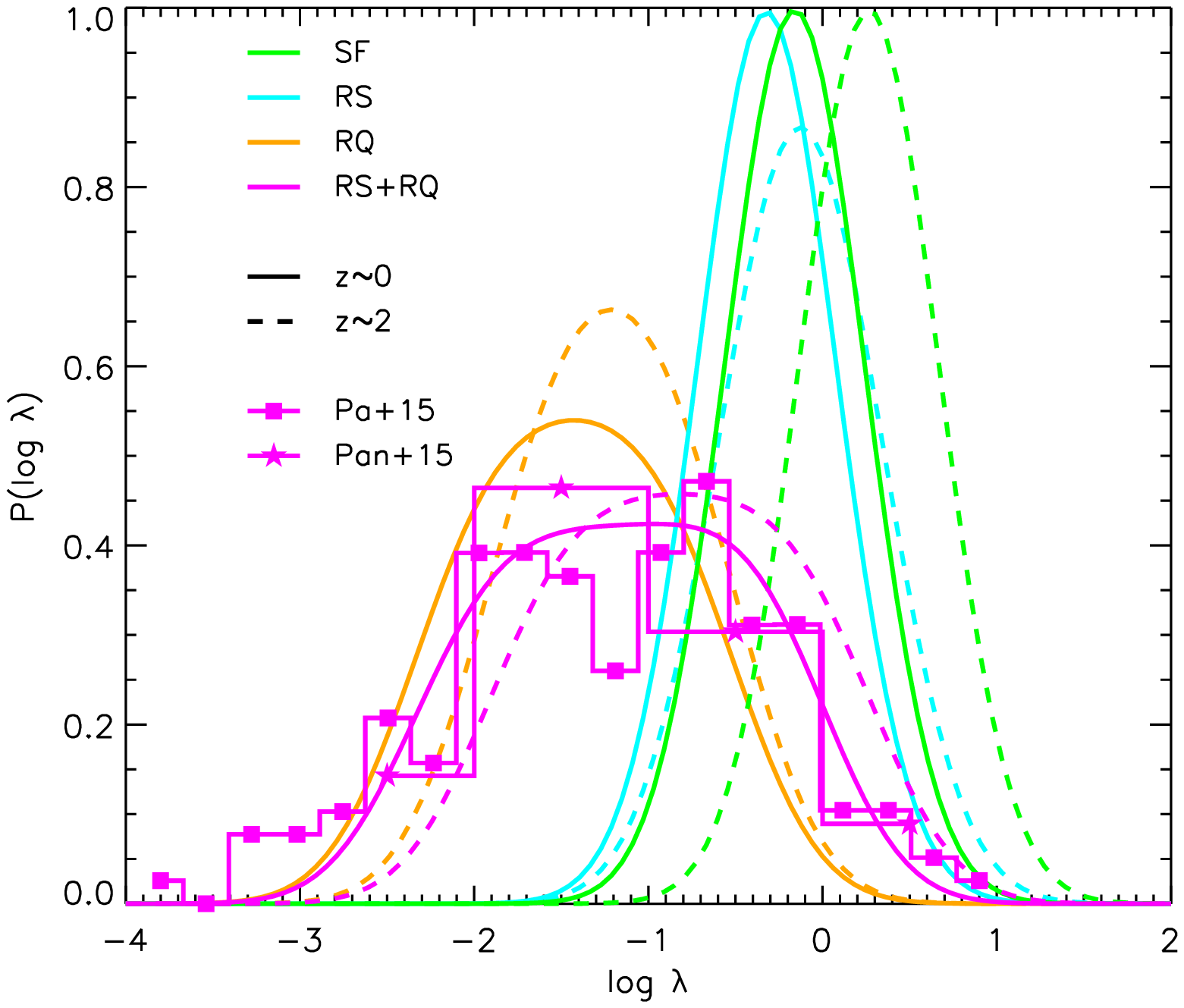}\caption{Schematic evolution with galactic age of the Eddington ratio $\lambda$, based on the star formation and BH accretion histories illustrated in Fig.~\ref{fig|timevo}. The curve is thin where the AGN is radio-silent, and thick where it is radio-active. The dots with arrows indicate the epochs when: the X-ray AGN luminosity exceeds $10^{42}$ erg s$^{-1}$; thin disk accretion sets in for $\lambda\la 0.3$. Colored strips as in Fig.~\ref{fig|timevo}. Bottom panel: corresponding Eddington ratio probability distribution (integral under curves normalized to unity) at redshift $z\sim 0$ (solid) and $2$ (dashed) for different populations: SFGs (green), RS AGNs (cyan), RQ AGNs (orange), RS+RQ AGNs (magenta). Data for RQ+RS AGNs at $z\sim 0$ and from Panessa et al. (2015; magenta stars) and at $z\sim 1-2$ from Padovani et al. (2015; magenta squares).}\label{fig|plambda}
\end{figure*}

The first is focused on the Eddington ratio distributions of the supermassive BHs hosted in SFGs, RQ and RL AGNs. In Fig.~\ref{fig|plambda} we illustrate the schematic evolution with galactic age of the Eddington ratio $\lambda$, based on the star formation and BH accretion histories presented in Fig.~\ref{fig|timevo}, and the corresponding $\lambda$-distributions. We expect the BHs hosted in SFGs to show a quiet narrow Eddington-ratio distribution centered around (mildly super-) Eddington values $\lambda\ga 1$. The same holds for radio-silent AGNs, with a distribution slightly offset toward smaller values of $\lambda$, that after the peak of BH activity starts decreasing. RQ AGNs are expected to feature a much broader distribution skewed toward $\lambda\la 0.3$, the value allowing the development of a thin-disk accretion suitable for nuclear radio emission (see Sect.~\ref{sec|radio}). Finally, low-$z$ steep-spectrum RL AGNs, featuring  low accretion rates, should feature a distribution shifted toward very small values $\lambda\la 0.01$; however, in the lack of a detailed physical understanding of their radio emission processes, we do not attempt definite predictions for this class.

The redshift dependence of the distributions for the different classes from $z\sim0$ to $2$ is mild, and dictated by the redshift evolution of the Eddington ratio in the early stages of the evolution, which is required by independent datasets on the AGN luminosity functions (see discussion in Sect.~\ref{sec|basics}). It is remarkable that our predictions for RQ AGNs are in agreement with the observational determinations, though within large uncertainties, by Panessa et al. (2015) and Padovani et al. (2015). This adds further validation to our overall picture for the radio emission from SFGs and RQ AGNs. A further, important but challenging test would consist in observationally determining the $\lambda$ distribution for pure SFGs hosting a radio-silent AGN (X-ray detected), and confront the outcome with our predictions for this population.

\begin{figure*}
\epsscale{0.6}\plotone{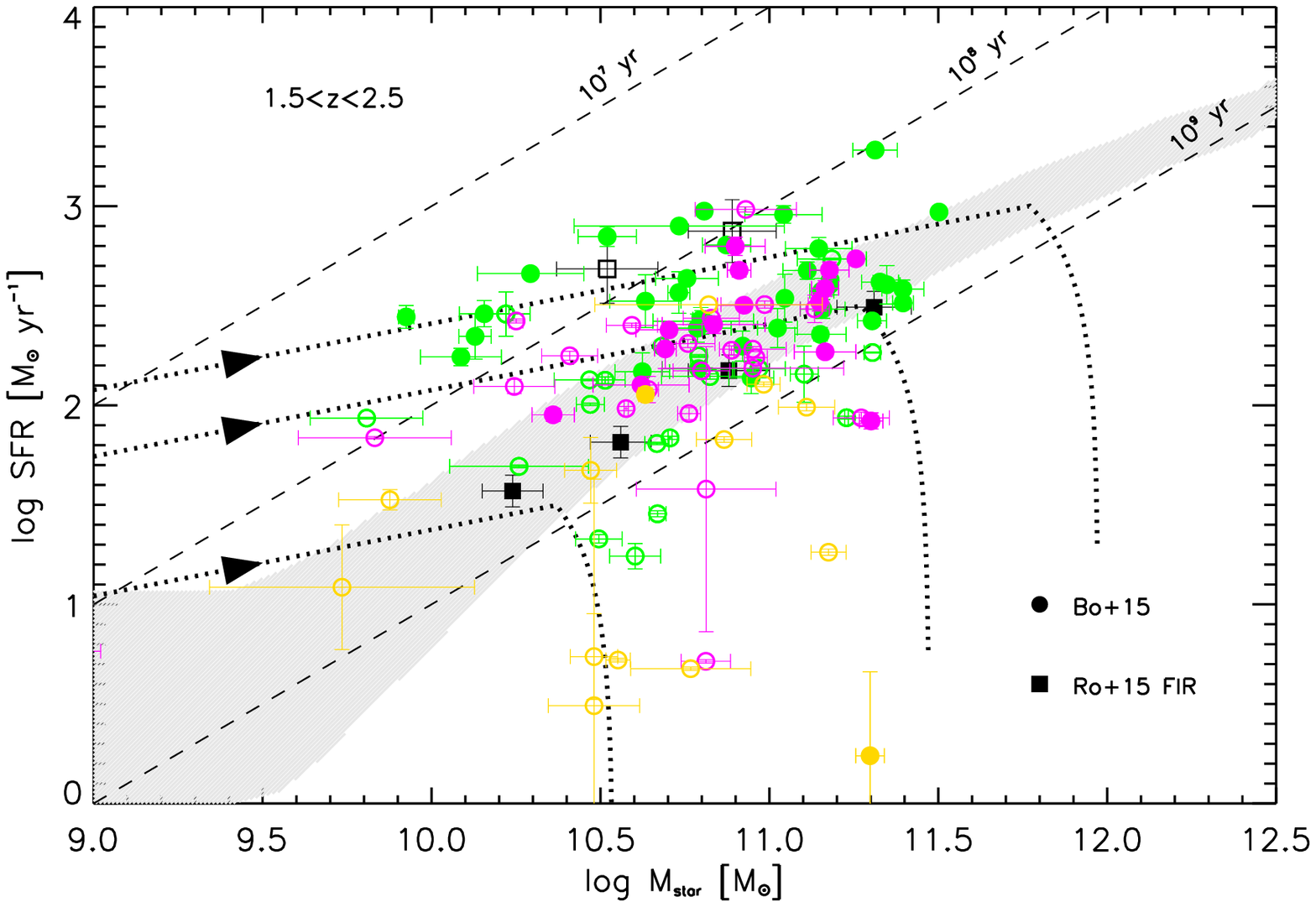}\\\epsscale{1.1}\plottwo{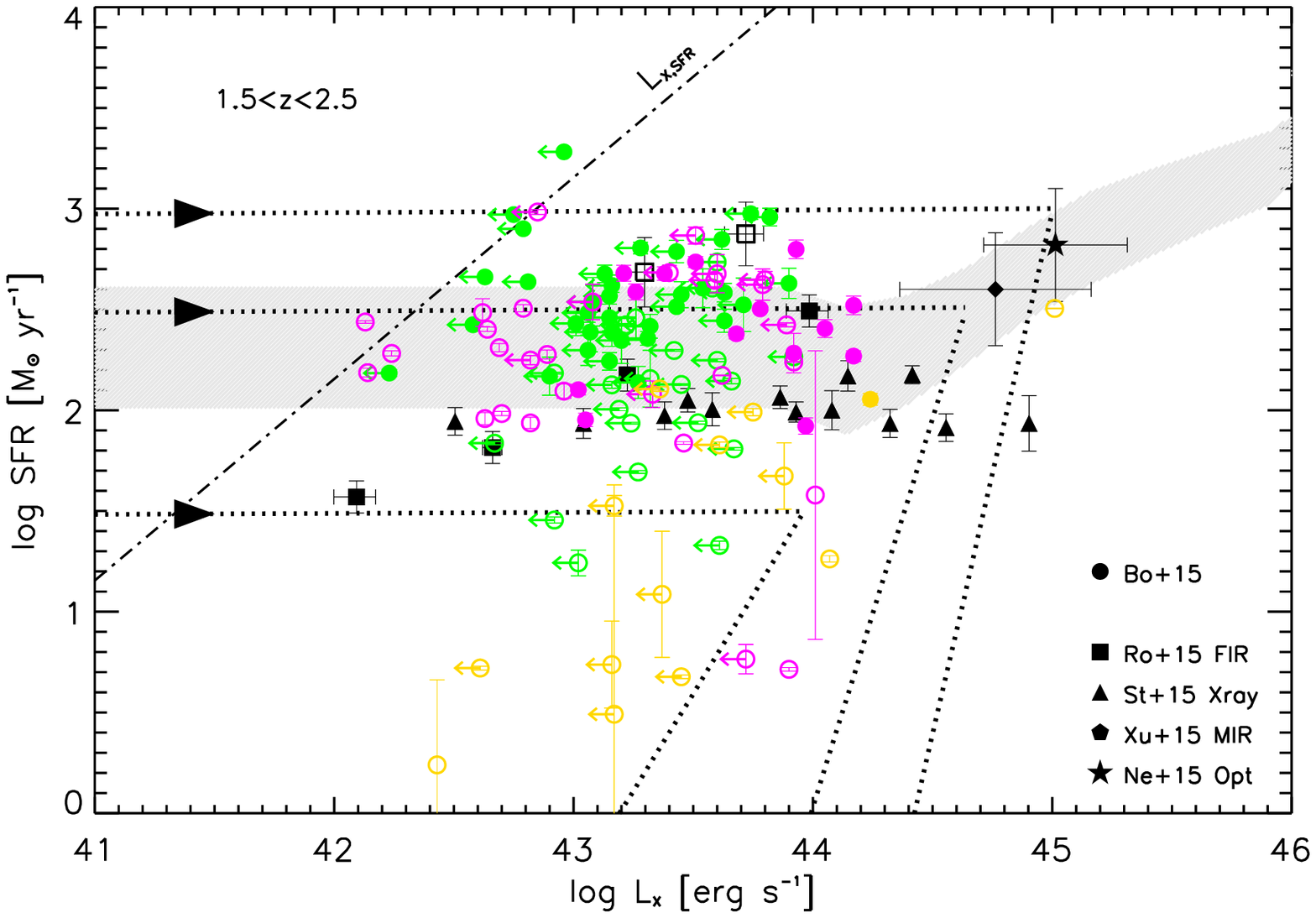}{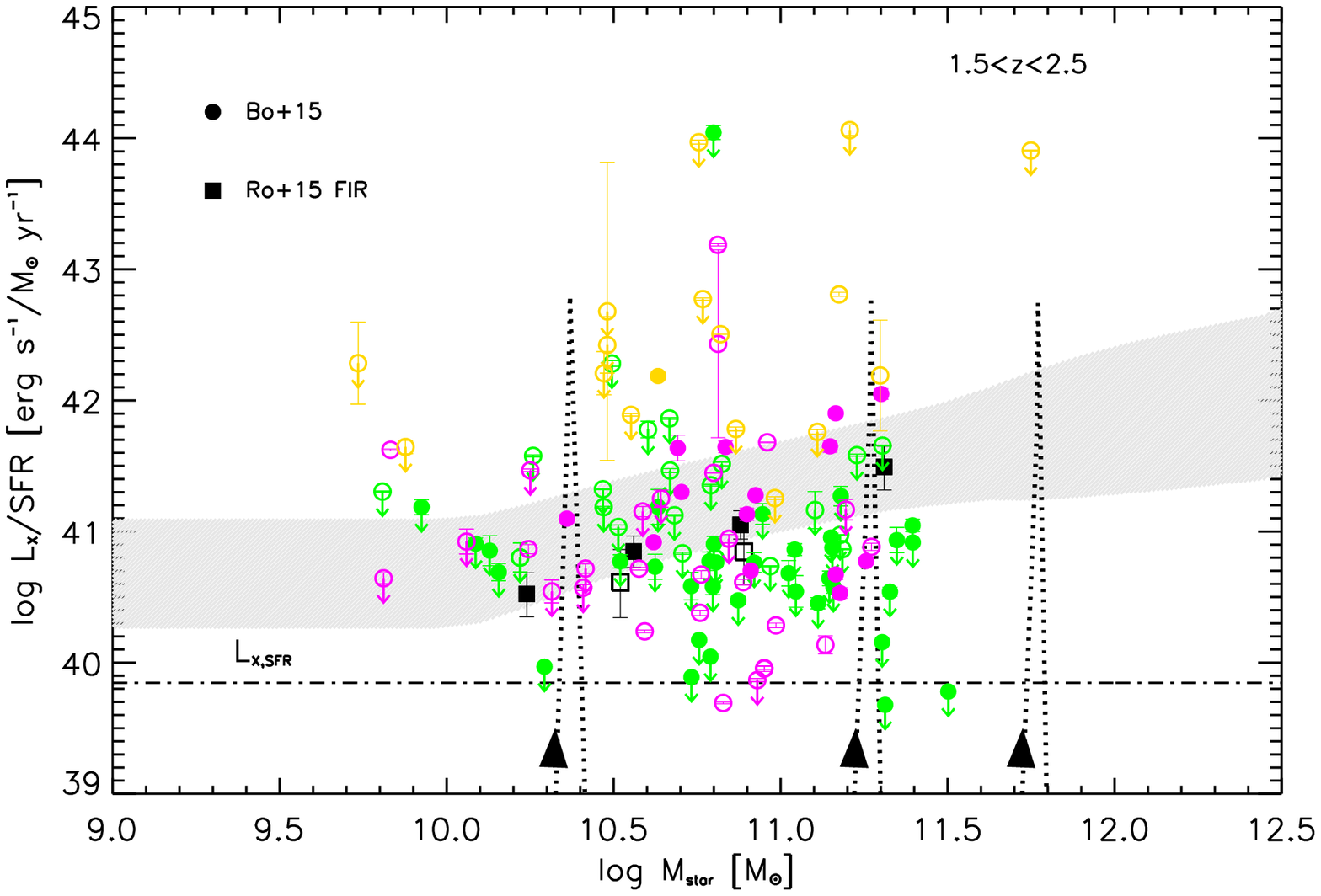}
\caption{Main sequences of SFGs and AGNs at $z\sim 2$: SFR vs. stellar mass (top panel), SFR vs. X-ray luminosity (bottom left) and ratio of X-ray luminosity to SFR vs. stellar mass (bottom right). Dotted lines illustrate three typical evolutionary tracks (forward time direction indicated by arrows) corresponding to peak values of the SFR $\dot M_\star\approx 30$, $300$, $1000\, M_\odot$ yr$^{-1}$, based on the star formation and BH accretion histories illustrated in Fig.~\ref{fig|timevo}. The grey shaded areas show the average relationships with their $2\sigma$ variance, computed as in Mancuso et al. (2016b) taking into account number density of galaxies and AGNs, and the relative time spent by individual objects in different portion of the track. Dashed lines highlight galaxy ages around $10^7$, $10^8$ and $10^9$ yr (from top to bottom). Dot-dashed lines show the X-ray luminosity expected from SFR according to the calibration by Vattakunnel et al. (2012). Data with radio information are from Bonzini et al. (2015; circles): green for SFGs (filled when SFR has been detected and hollow otherwise), magenta for RQ AGNs and yellow for RL AGNs (filled when SFR and X-ray luminosity have been detected and hollow otherwise). Other datasets are from Rodighiero et al. (2015; squares, filled for main-sequence objects and hollow for off-main sequence ones) for mass-selected galaxies, Stanley et al. (2015; triangles) for X-ray selected AGNs, Xu et al. (2015; pentagons) for mid-IR selected AGNs, and Netzer et al. (2015; stars) for optically selected quasars.}\label{fig|mainseq}
\end{figure*}

The second example is focused on the locus occupied by SFGs and AGNs on the main sequence diagrams: SFR vs. stellar mass, SFR vs. X-ray luminosity, and ratio of X-ray luminosity to SFR vs. stellar mass. In Fig.~\ref{fig|mainseq} we place in such diagrams at $z\sim 2$ the data with radio information by Padovani et al. (2015), highlighting the population of SFGs, RQ and RL AGNs. Other datasets referring to mass/far-IR selected galaxies (Rodighiero et al 2015), X-ray selected AGNs (Stanley et al. 2015), mid-IR selected AGNs (Xu et al. 2015) and optically selected quasars (Netzer et al. 2016) are also reported for completeness, although we caveat that the detection thresholds in SFR and X-ray luminosity are slightly different among them and with respect to the Padovani et al. sample.

In addition, we show three typical evolutionary tracks based on the star formation and BH accretion histories presented in Fig.~\ref{fig|timevo}, that correspond to values of SFR $\dot M_\star\approx 30$, $300$, and $1000\, M_\odot$ yr$^{-1}$ at the time when the AGN activity peaks. The shaded area shows the average relationships computed as in Mancuso et al. (2016b), taking into account the number density of galaxies and AGNs, and the relative time spent by individual objects in different portions of the evolutionary tracks. We expect galaxies without signs of nuclear activity to be rather young objects, featuring stellar masses appreciably smaller than implied by the average relationship at given SFR (corresponding to ages $\la$ few $10^8$ yr), and X-ray luminosities $L_X\la 10^{42}$ erg s$^{-1}$ mostly dominated by star formation. On the other hand, we expect RQ AGNs to be more evolved objects with stellar masses lying closer to the average relationship at given SFR (corresponding to ages $\la 10^9$ yr). They are also expected to host an X-ray detectable AGN with $L_X\ga 10^{42}$ erg s$^{-1}$, radio-silent when the star formation is still sustained, and progressively radio-active when star formation is in the way of getting quenched. Finally, low-$z$ steep spectrum RL AGNs are hosted mainly by galaxies in passive evolution so that their star formation activity is easily undetected at all.

Our expectations on SFGs and RQ AGNs are indeed consistent with the current multi-wavelength data with radio classification from Padovani et al. (2015). In some detail, most of the objects classified as SFGs have young ages $\la$ few $10^8$ yr and lie to the left of the average main sequence relationship at a given SFR; moreover, they have only upper limits on $L_X$ and on $L_X/$SFR ratios. Contrariwise, most of the objects classified as RQ AGNs have ages of several $10^8$ yr, stellar masses consistent with the average main sequence relationship, X-ray luminosities $L_X\ga 10^{42}$ erg s$^{-1}$ and $L_X/$SFR ratios appreciably higher than for SFGs.

At $z\la 0.6$ this picture has been partially confirmed over the large Herschel-ATLAS fields by Gurkan et al. (2015). However, at $z\ga 2$ the limited area around $0.3$ deg$^2$ of the Padovani et al. sample does not allow to populate with considerable statistics the diagrams at stellar masses $\ga 10^{11}\, M_\odot$ and X-ray luminosities $L_X\ga 10^{44}$ erg s$^{-1}$. We expect to find there an appreciable number of objects classified as RQ AGNs (but not of SFGs) with at least partially quenched SFR. It would be interesting to check such trend with multi-band data of comparable quality on larger areas, as it is the case for the Prandoni et al. (2017) WSRT observations over $6.6$ deg$^2$ in the framework of the LH Project, and as it will become routinely possible with the advent of the SKA and its precursors.

\begin{figure*}
\epsscale{1}\plotone{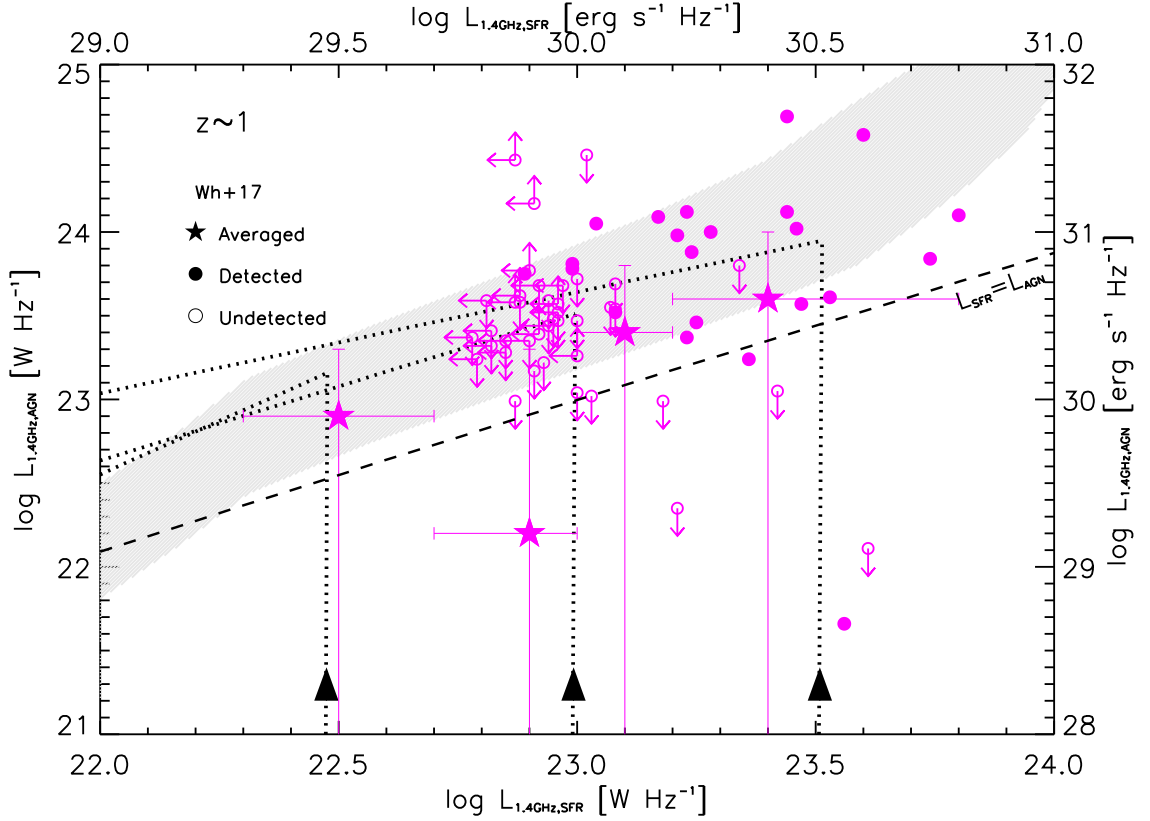}\caption{Radio luminosity at $1.4$ GHz from the central AGN vs. that from the star formation in the host, for radio-quiet systems at $z\sim 1$. Dotted lines illustrate three typical evolutionary tracks (forward time direction indicated by arrows) corresponding to peak bolometric AGN luminosities of $L_{\rm AGN}\approx 3\times 10^{45}$, $10^{46}$, and $3\times 10^{46}$ erg s$^{-1}$ (from left to right), based on the star formation and BH accretion histories illustrated in Fig.~\ref{fig|timevo}. The grey shaded areas show the average relationship with its $2\sigma$ variance, computed as in Mancuso et al. (2016b) taking into account number density of galaxies and AGNs, and the relative time spent by individual objects in different portion of the tracks. Dashed line shows the locus where the bolometric luminosities from the AGN and from star formation are equal. Data are from a sample of radio-quiet quasars at $z\sim 1$ by White et al. (2017): filled magenta circles refer to individual objects detected both in the far-IR and in the radio (error bars omitted for clarity); hollow magenta circles with arrows refer to undetected objects in the far-IR and/or in the radio; big stars with error bars refer to the median luminosities over the full sample.}\label{fig|RQ_decomp}
\end{figure*}

The third example is focused on disentangling the relative contribution from the active nucleus and from large-scale star formation to the radio emission of individual RQ AGNs. In Fig.~\ref{fig|RQ_decomp} we illustrate the locus occupied by RQ AGNs in a diagram where the radio luminosity $1.4$ GHz from the nucleus is plotted against that from star formation. First of all, we plot the data for RQ quasars at $z\sim 1$ by White et al. (2017); these authors followed-up an optically selected quasar sample with radio (FIRST) and far-IR (\textsl{Herschel}) observations to probe the relative contribution from the nucleus and from star formation to the radio emission.

As to our predictions, we show three typical evolutionary tracks based on the star formation and BH accretion histories presented in Fig.~\ref{fig|timevo}, that correspond to values of the peak AGN bolometric luminosities $L_{\rm AGN}\approx 3\times 10^{45}$, $10^{46}$, and $3\times 10^{46}$ erg s$^{-1}$, approximately the same range sampled by White et al. (2017). During the early stages of a galaxy's evolution, star formation is nearly constant while the AGN luminosity is exponentially increasing, to originate a vertical track in the diagram; after the AGN luminosity peaks, both the star formation and the AGN luminosity decrease, and the galaxy move to the left part of the diagram with a roughly flat track (the detailed shape depends on $\tau_{AGN}$, see Sect.~\ref{sec|timevo}). The shaded area shows the average relationship computed as in Mancuso et al. (2016b), taking into account the number density of AGNs with different luminosities and the relative time spent by individual objects in different portions of the evolutionary tracks.

Our expectations are in good agreement with the results from White et al. (2017) for RQ quasars, that tend to cluster close to the peaks of the individual evolutionary tracks, and lie within the average relationship (with its scatter) predicted by Mancuso et al. (2016b). All in all, for most of them the radio emission is found to be mainly contributed by the active nucleus. It would be interesting to test further our predictions (e.g., to independently constrain the timescale $\tau_{\rm AGN}$, see above) with larger samples spanning a wider luminosity range and attaining a higher sensitivity in the far-IR and radio bands.

\section{Summary}\label{sec|summary}

We have investigated the astrophysics of star-forming galaxies and radio active galactic nuclei (AGNs), and elucidated their statistical properties in the radio band including luminosity functions, redshift distributions, and number counts at sub-mJy flux levels, that will be crucially probed by next-generation radio continuum surveys.

We have achieved the goal following the model-independent approach by Mancuso et al. (2016a,b), based on two main ingredients: (i) the redshift-dependent star formation rate functions inferred from the latest UV/far-IR data from HST/Herschel, and related statistics of strong gravitationally lensed sources; (ii) deterministic tracks for the co-evolution of star formation and BH accretion in an individual galaxy, gauged on a wealth of multi-wavelength observations.

We have exploited such ingredients to compute the AGN duty cycle and probability of a SFG to host an AGN, so mapping the SFR functions into the observed bolometric AGN luminosity functions. Coupling these results with the radio emission properties associated to star formation and nuclear activity, we have computed relevant statistics at different radio frequencies, and disentangled the role of the SFGs and radio AGNs in different radio luminosity, radio flux, and redshift ranges.

Our main findings are the following:

$\bullet$ The local radio luminosity function of non RL sources is dominated by emission associated to star formation in galaxies below $10^{30}$ erg s$^{-1}$ Hz$^{-1}$, while above this value RQ AGNs powered by nuclear emission take over. At higher redshift the threshold separating the two contributions shifts toward brighter luminosities, up to $10^{32}$ erg s$^{-1}$ Hz$^{-1}$ at $z\la 2.5$. At any redshift, a fraction around $1\%$ of SFGs contains a growing nucleus, detectable in hard X rays but almost silent in the radio band. These conditions are expected to occur during the early stages of a massive galaxy's evolution when plenty of material is available for accretion onto the BH, enforcing a spherical, chaotic slim-disk accretion with negligible rotational energy available to fuel radio jets. On the other hand, RQ AGNs powered by nuclear emission are instead associated with a late stage of a massive galaxy's evolution when star formation is being quenched by AGN-driven outflows, while progressively lower accretion rates allow a standard thin disk to form around a spinning BH.

$\bullet$ At $1.4$ GHz SFGs and RQ AGNs starts to appreciably contribute to the counts at sub-mJy levels, progressively outnumbering the RL population. Around $S\sim 0.5$ mJy, a fraction of $1\%$ of the counts is contributed by strongly, lensed high-redshift SFGs. At fluxes below $\la 10^{-1}$ mJy SFGs dominates the total counts, and a fraction larger than $30\%$ of them will be high-redshift $z\ga 3$ sources. Interestingly, at these faint fluxes we expect to still find a $10\%$ contribution from RQ AGNs. These high-redshift systems will be particularly interesting targets for the next generation of ultradeep radio counts with SKA and its precursors.

$\bullet$ By comparing our results with the observations regarding the $1.4$ GHz luminosity function at different redshifts and counts, we do not find clear evidence for a substantial evolution in the normalization of the $L_{\rm radio}$ vs. SFR relationship, though the data are still consistent with a weak evolution. By the same token, we do not find up to $z\la 4$ signs of an appreciable decrease in synchrotron luminosity due to energy losses of relativistic electrons via Compton upscattering off CMB photons.

$\bullet$ At the low frequencies $\nu\la 150$ MHz currently explored by LOFAR and soon with the SKA, we have found that synchrotron self-absorption plays an important role in shaping the normalization of the euclidean part of the counts contributed from low-redshift SFGs. Comparing with current data, we have determined the average value $\nu_{\rm self}\approx 120$ MHz for the characteristic frequency of the synchrotron self-absorption process. Present observations with flux limit $\ga 1$ mJy probe the counts in a region dominated by RL sources; however, at subm-mJy flux levels, soon achievable with LOFAR and in full with the SKA, SFGs and RQ AGNs are expected to take over, with a substantial contribution from high-redshift $z\ga 3$ (unlensed) sources.

$\bullet$ At higher frequencies $\nu\ga 10$ GHz to be probed with the SKA and its precursors, the behavior in terms of the counts is similar. The emission from SFGs is increasingly dominated by free-free at sub-mJy levels, and at $z\ga 1$ in terms of redshifts. This makes $\nu\approx 10$ GHz an optimal frequency to study the free-free emission from SFGs over an extended redshift range $z\sim 1-8$, since there it dominates over synchrotron and it is redshifted to restframe frequencies appreciably below $\nu\approx 150$ GHz where dust emission starts to become relevant. Moreover, given that the synchrotron emission at high redshift $z\ga 6$ may be considerably affected by inverse Compton scattering off the CMB photons, high frequency surveys at $\nu\ga 10$ GHz can be extremely useful to search for high-$z$ SFGs via their free-free emission.

$\bullet$ We have highlighted that substantially different Eddington ratio distributions and different positions on the main sequence diagrams are expected for SFGs, RQ and RL AGNs. With respect to SFGs, radio AGNs are expected to be older systems, with higher stellar masses at given SFR, higher X-ray nuclear luminosity and $L_X/$SFR ratios, and broader distributions of Eddington ratio skewed toward lower values $\lambda\la 0.3$. Finally, an appreciable fraction of RQ systems with X-ray luminosities $\ga 10^{43}$ erg s$^{-1}$ should feature already suppressed SFR with respect to objects classified as radio-emitting SFGs. Optically-selected, radio-quiet quasars are indeed found to have their radio emission mostly contributed by the active nucleus. Testing effectively these predictions requires data with multi-band coverage (X-ray, radio, IR) on large areas $\ga$ several deg$^2$, as it will become routinely possible with the advent of the SKA and its precursors.

\begin{acknowledgements}
We are grateful to M. Massardi, P. Padovani, and S.V. White for stimulating discussions. We acknowledge the referee for a constructive report. Work partially supported by PRIN INAF 2014 `Probing the AGN/galaxy co-evolution through ultra-deep and ultra-high-resolution radio surveys' and by PRIN MIUR 2015 `Cosmology and Fundamental Physics: illuminating the Dark Universe with Euclid'. AL and FP acknowledge the RADIOFOREGROUNDS grant (COMPET-05-2015, agreement number 687312) of the European Union Horizon 2020 research and innovation programme. JGN acknowledges financial support from the Spanish MINECO for a 'Ramon y Cajal' fellowship (RYC-2013-13256) and the I+D 2015 project AYA2015-65887-P (MINECO/FEDER).
\end{acknowledgements}

\clearpage

\clearpage

\clearpage

\clearpage

\clearpage

\clearpage

\clearpage

\clearpage

\clearpage

\clearpage

\clearpage

\clearpage

\clearpage

\clearpage

\clearpage

\clearpage

\clearpage
\begin{deluxetable}{lcccccccccccccccc}
\tabletypesize{\scriptsize}\tablewidth{0pt}\tablecaption{SFR Function
Parameters}\tablehead{\colhead{Parameter} & &
\colhead{$p_0$} & & \colhead{$p_1$} & & \colhead{$p_2$} & &
\colhead{$p_3$}} \startdata
$\log \mathcal{N}(z)$ & &$-2.13$& &$-8.90$& &$18.07$& &$-9.58$\\
$\log\dot M_{\star, c}(z)$ & & $0.72$& &$8.56$& &$-10.08$& &$2.54$\\
$\alpha(z)$& &$1.12$& &$3.73$& &$-7.80$& &$5.15$\\
\enddata
\tablecomments{We have adopted the Meurer-Calzetti law to compute dust correction for UV data, and the Clemens et al. (2013) prescriptions to subtract cirrus emission from low-redshift $z\la 1$ far-IR data.}
\end{deluxetable}

\end{document}